\documentclass[preprint,12pt]{article}
\usepackage{authblk}

\usepackage{braket}
\usepackage[square, sort&compress, numbers]{natbib}
\usepackage{color}
\usepackage{amssymb,amsmath,amsmath,amsfonts,amsthm,bm}
\usepackage{hyperref}
\usepackage{xcolor}
\hypersetup{
    colorlinks,
    linkcolor={red!75!black},
    citecolor={blue!75!black},
    urlcolor={blue!75!black}}
\usepackage[utf8]{inputenc}
\usepackage{physics}
\usepackage{graphicx}
\usepackage{caption}
\usepackage{subcaption}
\usepackage{tabularx}
\usepackage{enumitem}
\usepackage{cleveref}
\usepackage{threeparttable}
\usepackage{lscape}
\usepackage{natbib}

\usepackage{flushend}
\usepackage{color}
\usepackage{float}
\usepackage{multirow}

\definecolor{red}{rgb}{1.0,0.0,0.0}

\begin{document}
\title{\bf{Exploring non-radial oscillation modes in dark matter admixed neutron stars}} 

\author[a]{Pratik Thakur\thanks{\texttt{thakur.16@iitj.ac.in}}}
\author[a]{Anil Kumar\thanks{\texttt{anil.1@iitj.ac.in}}}
\author[b]{Vivek Baruah Thapa\thanks{\texttt{thapa.1@iitj.ac.in}}}
\author[c]{Vishal Parmar\thanks{\texttt{vishalparmar@iitj.ac.in}}}
\author[a]{Monika Sinha\thanks{corresponding author: \texttt{ms@iitj.ac.in}}}

\affil[a]{Indian Institute of Technology Jodhpur, \\Jodhpur 342037, India}
\affil[b]{Department of Physics, Bhawanipur Anchalik College,\\Barpeta, Assam 781352, India}
\affil[c]{INFN, Sezione di Pisa, Largo B. Pontecorvo 3, I-56127 Pisa, Italy}
\maketitle
\begin{abstract}
Because of their extreme densities and consequently, gravitational potential, compact objects such as neutron stars can prove to be excellent captors of dark matter particles. Considering purely gravitational interactions between dark and hadronic matter, we construct dark matter admixed stars composed of two-fluid matter subject to current astrophysical constraints on maximum mass and tidal deformability. We choose a wide range of parameters to construct the dark matter equation of state, and the DDME2 parameterization for the hadronic equation of state. We then examine the effect of dark matter on the stellar structure, tidal deformability and non-radial modes considering the relativistic Cowling approximation. We find the effect on $p$-modes is substantial, with frequencies decreasing up to the typical $f\text{-}$mode frequency range for most stars with a dark matter halo. The effects on the $f\text{-}$mode frequency are less extreme. Finally, we find the most probable values of the dark matter parameters that satisfy the observational constraints.

\end{abstract}
\newpage
\section{Introduction}\label{intro}
There are indirect pieces of evidence from cosmological and astrophysical observations such as rotation curves of individual galaxies, observed galaxy clusters, and anisotropic microwave background that the dominant contribution to the matter in the universe is in the form of ``dark matter" (DM) whose cross-section with baryons is very small \citep{2005PhR...405..279B}. The search for direct detection of DM particles is also ongoing in several experiments \cite{aalbers2016darwin,PhysRevLett.109.021301,alekhin2016facility}. A well-studied model for DM involves weakly interacting massive particles (WIMPs) with masses ranging from the $\mathrm{GeV}$ to the TeV scale. The upper bound of WIMPs scattering off the nuclei is estimated to be $\sim 10^{-47}$ cm$^2$ from underground direct search experiments such as CDMS \cite{2010Sci...327.1619C}, Xenon1T \cite{2017PhRvL.119r1301A} and XENONnT \cite{PhysRevLett.131.041003}. Despite the tiny value of the scattering cross-section of DM with ordinary matter, the gravitational impact of DM is very significant. However, nearly all properties of DM remain largely unknown. If DM exists, understanding its nature would provide significant advancements in our knowledge beyond standard model physics \cite{2021PrPNP.11603824K}. 

Many cosmological problems such as the "core-cusp problem," the "missing satellite problem", and the "too big to fail problem" could potentially be addressed by considering the DM as self-interacting \cite{2017PhRvD..96b3005M,2015PhRvD..92f3526K}. Self-interacting DM particles may be self-annihilating or non-self-annihilating. If they are self-annihilating, after their capture by compact objects, they will annihilate, releasing heat inside the compact objects \cite{2010PhRvD..81l3521D,2024JCAP...04..006B,2018JHEP...08..069C,2021PhRvD.103d3019G}. On the other hand, if they do not annihilate, these particles will accumulate within compact objects, provided the dark matter particle is a WIMP \cite{PhysRevD.93.023508,2010PhRvD..82f3531K}. Hence, the compact star with a large number of accreted DM in the core will easily exceed the Schwarzschild limit and consequently collapse. The asymmetric DM is an interesting alternative of the annihilating WIMP paradigm \cite{nussinov1985technocosmology,BARR1990387,2006PhRvD..74i5008G,2009PhRvD..80c7702F,2007PhRvD..76j5004R,2009arXiv0911.0931S,PhysRevD.78.115010,PhysRevD.79.015016,PhysRevD.81.097704,2012JCAP...03..019M,2011JHEP...09..128F,2013JCAP...01..021G,2012NuPhB.854..666A,PhysRevLett.108.011301,PhysRevD.85.014504,PhysRevD.84.096008,2011JHEP...10..110G,PhysRevD.84.123505,PhysRevD.84.035007}.

Because of the gravitational impact of DM, its properties can be constrained from its effect on the formation and evolution of stars \cite{2009ApJ...705..135C,2009MNRAS.394...82S}.  Compact objects, in this sense, are potential candidates to study DM, not only because of immense gravity which increases the possibility of the capture of DM particles, but also for their high baryonic density, which increases the probability of DM-nucleon scattering \cite{2008PhRvD..77b3006K,2010PhRvD..81l3521D,2010PhRvD..81h3520M,2010PhRvD..82f3531K,2015PhRvL.115k1301B,2014JCAP...05..013G,bramanteDarkMatterCompact2023}. DM admixed compact stars have been explored in several previous works, considering a variety of DM and baryonic matter EOSs \cite{2021ApJ...914..138C,2015ApJ...812..110L,2011PhLB..695...19C,PhysRevD.84.107301,2017ApJ...835...33R,PhysRevD.97.123007,PhysRevD.99.083008,PhysRevD.99.063015,Nelson_2019,10.1093/mnras/staa1435,2022EPJC...82..366Z,ben_kain,mukhopadhyayQuarkStarsAdmixed2016,kwing_leung,shirkeRmodesNewProbe2023,2024arXiv240318740S}.

The recently found lower bound in maximum attainable mass of compact stars \cite{2010Natur.467.1081D} raises questions on the appearance of heavier strange and non-strange baryons in the inner core of the baryonic stars \cite{2017hspp.confj1002B}. However, the appearance of these degrees of freedom softens the matter at high density regimes which lowers the maximum attainable mass. If by altering the theoretical proposed model for this kind of matter- for example by taking density-dependent interactions among baryons \cite{2010PhRvC..81a5803T,2011PhRvC..84e4309R,2020PhLB..80035065T}, the matter becomes stiff, the maximum limit of tidal deformability can not be fulfilled \cite{2021MNRAS.507.2991T}. However, as will be shown, the presence of DM can significantly alter the stellar structure, and both the lower limit of maximum attainable mass and the upper limit of tidal deformability can be achieved simultaneously. Consequently, the study of DM admixed compact star property which is very important from particle physics as well as from astrophysical aspects.

We have constructed the DM admixed compact star considering only the gravitational interaction of self-interacting but non-self-annihilating DM with ordinary matter as in refs. \cite{2014PhRvC..89b5803X,arpan_das}. Interested readers may refer to \cite{2009APh....32..278S,2011PhLB..695...19C,ben_kain}. Due to the lack of evidence and proper knowledge of DM interaction with ordinary matter, we avoid such beyond-standard model interactions in the present work. For a compact star, we consider a neutron star (NS) to be composed of nucleonic matter or hyperons are also present at high density. We study the effect on the NS properties connected to the recent and upcoming stellar observations considering the scenario in DM admixed NS. Further, in this context, we will discuss the properties of non-radial oscillations for such stars. 

There are several quasi-normal modes like the fundamental ($f$)-mode, pressure ($p$)-modes, gravity ($g$)-modes, spacetime ($w$)-mode, etc \citep{1999LRR.....2....2K,lindblom1983quadrupole,detweiler1985nonradial}, each classified based on the restoring forces which work bring the star to equilibrium. For example, the $f\text{-}$ and $p_1\text{-}$ (with only one radial node in fluid perturbation amplitude) modes, which are acoustic waves in the star, are restored by fluid pressure while $g-$modes which arise due to density discontinuities or temperature and composition variations are restored by gravity (buoyancy). Several of these modes can be excited during a supernova explosion or in a remnant of a binary merger \cite{2001MNRAS.320..307K,2011MNRAS.418..427S,2020PhRvD.101h4039V}. Even during the inspiral phase of compact stars, the $f\text{-}$mode may be excited \cite{2017ApJ...837...67C}. Quadrupolar oscillations ($l=2$) of all modes couple to gravitational waves. With the advent of enhanced next-generation telescopes like the Cosmic Explorer and the Einstein telescope which carry about 10 times the sensitivity of Advanced LIGO, the possibility of detection of these modes increases \cite{2001MNRAS.320..307K}.

\section{Hadronic matter model}

 The model for hadronic matter (HM) we discussed here, we assume that normal matter is composed of protons, neutrons, hyperons ($\Lambda, \Sigma$ and $\Xi$), and electrons. We consider the relativistic mean field model for ordinary HM where the interaction between these baryons is carried by isoscalar-scalar $\sigma$, isoscalar-vector $\omega$, and isovector-vector $\rho$ mesons along with density-dependent interaction. Additionally, the interaction between hyperons is mediated through isoscalar-scalar $\sigma^*$ and isoscalar-vector $\phi$ meson.The Lagrangian density for HM is given as \cite{1996cost.book.....G}
\begin{align}
\mathcal{L}_{HM} &= \sum_{b} \bar{\psi}_b(i\gamma_{\mu} D^{\mu} - m^{*}_b) \psi_b + \frac{1}{2}(\partial_{\mu}\sigma\partial^{\mu}\sigma - m_{\sigma}^2{\sigma}^2) + \frac{1}{2}(\partial_{\mu}\sigma^*\partial^{\mu}\sigma^* - m_{\sigma^*}^2 {\sigma^*}^2)\nonumber\\ 
&-  \frac{1}{4}\omega_{\mu\nu}\omega^{\mu\nu} + \frac{1}{2}m_{\omega}^2\omega_{\mu}\omega^{\mu} - \frac{1}{4}\phi_{\mu\nu}\phi^{\mu\nu} + \frac{1}{2}m_{\phi}^2\phi_{\mu}\phi^{\mu} \color{black}{- \frac{1}{4}\boldsymbol{\rho}_{\mu\nu}\cdot\boldsymbol{\rho}^{\mu\nu}}
+\frac{1}{2}{m_{\rho}^2}\boldsymbol{{\rho}_{\mu}}\cdot\boldsymbol{{\rho}^{\mu}}
\end{align}
Here $b$ denotes baryons, and covariant derivative $D_{\mu} = \partial_\mu + ig_{\omega b} \omega_\mu + ig_{\phi b} \phi_\mu + ig_{\rho N} \boldsymbol{\tau}_{N3} \cdot \boldsymbol{\rho}_{\mu}$. $\psi_b$ is the baryonic wavefunction, $\sigma$ the $\sigma$-meson, $\sigma^*$ the $\sigma^*$-meson, $\omega_\mu$ the $\omega$-meson, $\phi_\mu$ the $\phi$-meson and $\rho_\mu$ the $\rho$-meson fields. The effective baryon mass is ${{m_b}^*}={m_b}-g_{\sigma{b}}\sigma - g_{\sigma^*{b}}\sigma^*$. The antisymmetric field terms due to vector meson fields are given by $\omega_{\mu \nu} = \partial_{\mu}\omega_{\nu} - \partial_{\nu}\omega_{\mu}$, $\phi_{\mu \nu} = \partial_{\mu}\phi_{\nu} - \partial_{\nu}\phi_{\mu}$ and $\boldsymbol{\rho}_{\mu \nu} = \partial_{\mu} \boldsymbol{\rho}_{\nu} - \partial_{\nu}\boldsymbol{\rho}_{\mu}$.  The isoscalar meson-baryon couplings vary with density as
\begin{equation}\label{eqn.dd_isoscalar}
g_{i b}(n)= g_{i b}(n_{0}) f_i(x) \quad \quad \text{for }i=\sigma,\sigma^*,\omega,\phi
\end{equation}
where the function is given by
\begin{equation}\label{eqn.func}
f_i(x)= a_i \frac{1+b_i (x+d_i)^2}{1+c_i (x +d_i)^2}
\end{equation}
with $x=n/n_0$ and $a_i$, $b_i$, $c_i$, $d_i$ the parameters that describe the density-dependent nature of the coupling parameters. The isovector-vector $\rho$-meson coupling with baryons is given by $g_{\rho b}(n)= g_{\rho b}(n_{0}) e^{-a_{\rho}(x-1)}$.\\

With this model, the energy density $\varepsilon$ and pressure $p$ of hadronic matter are given by
\begin{align}
\varepsilon_{HM} &= \frac{1}{2} m_{\sigma}^2 \sigma^2 + \frac{1}{2} m_{\sigma^*}^2 {\sigma^*}^2 + \frac{1}{2} m_{\omega}^2 \omega_{0}^2 + \frac{1}{2} m_{\phi}^2 \phi_{0}^2 + \frac{1}{2} m_{\rho}^2 \rho_{03}^2 + \frac{1}{4\pi^2} \sum_b \bigg[ p_{F_b} E_{F_b}^3 - \frac{m_{b}^{*2}}{2} \bigg( p_{F_b} E_{F_b}\nonumber\\
&\quad + m_{b}^{*2} \ln \left( \frac{p_{F_b} + E_{F_b}}{m_{b}^{*}} \right)\bigg) \bigg] + \frac{1}{4\pi^2} \sum_l \left[ p_{F_l} E_{F_l}^3 - \frac{m_l^2}{2} \left( p_{F_l} E_{F_l} + m_l^2 \ln \left( \frac{p_{F_l} + E_{F_l}}{m_l} \right) \right) \right]
\end{align}

and,
\begin{align}
p_{HM} & = -\frac{1}{2}m_{\sigma}^2 \sigma^{2} + \frac{1}{2}m_{\sigma^*}^2 {\sigma^*}^{2} + \frac{1}{2} m_{\omega}^2 \omega_{0}^2 + \frac{1}{2} m_{\phi}^2 \phi_{0}^2 + \frac{1}{2}m_{\rho}^2 \rho_{03}^2 + \sum_b \frac{1}{12\pi^2} \bigg[ p_{{F}_b} E^3_{F_b} - \frac{m_{b}^{*2}}{2} \bigg( 5 p_{{F}_b} E_{F_b}\nonumber\\& \quad - 3 m_{b}^{*2} \ln \left( \frac{p_{{F}_b} + E_{F_b}}{m_{b}^{*}} \right)\bigg)\bigg] + \frac{1}{12 \pi^2}\sum_l \left[ p_{{F}_l} E^3_{F_l} - \frac{m_{l}^{2}}{2} \left( 5 p_{{F}_l} E_{F_l} - 3 m_{l}^{2} \ln \left( \frac{p_{{F}_l} + E_{F_l}}{m_{l}} \right) \right) \right]\nonumber\\ & 
\quad + n\Sigma^r 
\end{align}
respectively with $p_{F_j}$, $E_{F_j}$ denoting the Fermi momentum and Fermi energy of the $j-$th fermion in the system.
Here the rearrangement term $\Sigma^{r}$ arises due to the density dependence of coupling parameters (to maintain thermodynamic consistency) and is given by 
\citep{2001PhRvC..64b5804H}  
\begin{equation}
\begin{aligned}
\Sigma^{r} & = \sum_{b} \left[ \frac{\partial g_{\omega b}}{\partial n}\omega_{0}n_{b} +\frac{\partial g_{\phi b}}{\partial n}\phi_{0}n_{b} - \frac{\partial g_{\sigma b}}{\partial n} \sigma n_{b}^s - \frac{\partial g_{\sigma^* b}}{\partial n} \sigma^* n_{b}^s + \frac{\partial g_{\rho b}}{\partial n} \rho_{03} \boldsymbol{\tau}_{N3} n_{b} \right].
\end{aligned}
\end{equation} 
The parameters at nuclear saturation density ($n_0$), saturation energy ($E_0$), incompressibility ($K_0$), skewness ($Q_0$), effective nucleon mass ($m^*_N/m_N$), symmetry energy ($E_{sym}$) and its slope $L_{sym}$ are given in table \ref{Tab:nuclear_param}.
\begin{table}[h!]
\begin{center}
\begin{tabular}{lllllll}
\hline
$n_0$ (${fm^{-3}}$) & $E_0$ (MeV) & $K_0$ (MeV)& $Q_0$ (MeV)& $E_{sym}$ (MeV)& $L_{sym}$ (MeV)& $m^*_N/m_N$ \\
\hline
0.152  & -16.14 & 250.9 & 478.9 & 32.3  & 51.3 & 0.57
\\
\hline
\end{tabular}
\caption{The nuclear parameters at $n_0$}.
\label{Tab:nuclear_param}
\end{center}
\end{table}
\section{Dark matter model} 
Because of the unclear picture of the DM mass and interactions, theoretically many models of the DM are there that can be incorporated into viable beyond standard model theories. This matter can be fermionic or bosonic, self-interacting via attractive or repulsive potential or even non-interacting \cite{2017PhRvD..96b3005M}. Following previous works \cite{arpan_das,shirkeRmodesNewProbe2023}, we construct the DM model in the relativistic mean-field approach, considering `dark' hadrons that self-interact via `dark scalar' and `dark vector' bosons. We consider fermionic DM with cases of no self-interactions, either scalar or vector self-interactions, or both. DM is assumed to interact with HM only gravitationally. The Lagrangian of DM is given by \cite{arpan_das}: 

\begin{align}
    \mathcal{L}_{DM} = &\bar{\psi_D}\qty[\gamma_\mu\qty(i\partial^\mu-g_{vd}V^\mu)-\qty(m_D-g_{sd}\phi_D)]\psi_D+\frac{1}{2}\qty(\partial_\mu\phi_D\partial^\mu\phi_D-m_{sd}^2\phi_D^2)\nonumber\\
    &-\frac{1}{4}V_{\mu\nu,D}V^{\mu\nu}_D +\frac{1}{2}m_{vd}^2V_{\mu,D}V^{\mu}_D
\end{align}

Here $\psi_D$, $\phi_D$, and $V^\mu_D$ represent the fermionic DM, `dark scalar meson', and `dark vector meson' respectively. $m_D$ is the bare mass of the fermionic DM. $m_{sd}$ and $m_{vd}$ are the corresponding masses of the scalar and vector `dark mesons' respectively.
$g_{sd}$ and $g_{vd}$ are the corresponding coupling strengths. From this Lagrangian, the pressure, energy density, and number density of DM can be given as \cite{arpan_das,mukhopadhyayQuarkStarsAdmixed2016,shirkeRmodesNewProbe2023,ben_kain,panotopoulosDarkMatterEffect2017}

\begin{equation}
\varepsilon_{DM} =\frac{1}{\pi^2} \left[ k_{{F}_D} E^3_{F_D} - \frac{m_{D}^{*2}}{8} \left\{ k_{{F}_D} E_{F_D} + m_{D}^{*2} \ln \left( \frac{k_{{F}_D} + E_{F_D}}{m_{D}^{*}} \right) \right\} \right]+ \frac{n_{DM}^2c_v^2}{2}+ \frac{m_{sd}^2\phi_{D_0}^2}{2},
\end{equation}

\begin{equation}
\begin{aligned}
p_{DM} &= \frac{1}{12\pi^2} \left[ k_{{F}_D} E^3_{F_D} - \frac{m_{D}^{*2}}{2} \left\{ 5 k_{{F}_D} E_{F_D} - 3 m_{D}^{*2} \ln \left( \frac{k_{{F}_D} + E_{F_D}}{m_{D}^{*}} \right) \right\} \right] + \frac{n_{DM}^2c_v^2}{2} - \frac{m_{sd}^2\phi_{D_0}^2}{2},
\end{aligned}
\end{equation}



\begin{equation}
   n_{DM} = \frac{k_{F_D}^3}{3\pi^2}
\end{equation}

where $k_{F_D}$ is Fermi momentum of DM particle and $E_{F_D}=\sqrt{k_{F_D}^2+ m_{D}^{*2}}$ its Fermi energy.
$m_D^*$ is the effective DM fermion mass, given by
\begin{equation}
    m_D^* = m_D - g_{sd}\phi_{D_0}
\end{equation}
and $\phi_{D_0}$ is \cite{panotopoulosDarkMatterEffect2017}
\begin{equation}
    \phi_{D_0} = \frac{g_{sd}}{m_{sd}^2}\langle\bar\psi_D\psi_D\rangle =\frac{c_{s}^2}{g_{sd}}\times \frac{{m_D^*}}{2\pi^2}\qty[k_{F_D}E_{F_D} - {m^{*2}_D}\ln\left( \frac{k_{{F}_D} + E_{F_D}}{m_D^*} \right)]
\end{equation}


where $c_v= g_{vd}/m_{vd}$ and $c_s= g_{sd}/m_{sd}$ are the scalar and vector free parameters. 

\subsection{DM parameters} \label{sec: DM params}

In this work, we construct DM EOSs by varying several free parameters- the free DM particle mass, $m_D$, the scalar and vector free parameters, $c_s$, and $c_v$. Since these parameters can take on arbitrary values, we limit them as follows: 
\begin{itemize}
    \item $m_D$ is taken in the range of $0.2-2.0$ $\mathrm{GeV}$ in intervals of $0.1$ $\mathrm{GeV}$. This range of DM particle masses is arbitrary but gives pure DM stars with mass in the order of solar mass \cite{kwing_leung,liu_wei_DM_effects_2024}.
    \item Ranges of $c_s$ and $c_v$ are arbitrary. Although \cite{arpan_das} explores $c_s$ and $c_v$ in the ranges of $1-7$ $\mathrm{GeV}^{-1}$ and $10-14$ $\mathrm{GeV}^{-1}$ respectively, we vary both of them from $0-50$ $\mathrm{GeV}^{-1}$ in steps of $2$ $\mathrm{GeV}^{-1}$. With hyperons we choose the value of $c_v$ up to $100$ $\mathrm{GeV}^{-1}$.
    \item Not all combinations of $c_s$, $c_v$ and $m_D$ give valid EOSs. We discard EOSs where the ${\mathrm{d}p}/{\mathrm{d}\varepsilon}$ becomes negative or where pressure itself becomes negative.
    \item As will be explained in section \ref{sec: DM param TOV}, to obtain the DM EOS, we need to fix another parameter- the total DM mass fraction, $M_f$. This is the ratio of the mass of DM in a star to the total mass of the star. We vary this parameter from $1\%-20\%$ in steps of $1\%$
\end{itemize}

\section{Stellar Structure and Tidal Deformability}\label{stellar structure eqns}

\subsection{TOV equations}

In our model, DM interacts with HM of the star only via gravitational interaction. Hence, the stellar structure of DM admixed NSs will be governed by the two-fluid stellar structure equations. The two fluids in our case are the HM and the DM. The equilibrium configurations of non-rotating relativistic stars are obtained by solving the Tolmann-Oppenheimer-Volkoff (TOV) equations with the spherically symmetric line element
\begin{equation}
    ds^2= -\mathrm{e}^{2\Phi}\ dt^2+ \mathrm{e}^{2\Lambda}\ dr^2+r^2\ d\theta^2+r^2\sin^2\theta\ d\phi^2\label{metric}
\end{equation}

The single fluid TOV equations are
\begin{equation}
    \begin{aligned}
        \frac{\dd m}{\dd r} &= 4\pi r^2 \varepsilon\\[5pt]
        \frac{\dd \Phi}{\dd r} &=\frac{m+4 \pi r^3p}{r (r- 2 m)}\\[5pt]
        \frac{\dd p}{\dd r} &=-(p+\varepsilon)\frac{\dd \Phi}{\dd r}\\[5pt]
        \Lambda (r)&= -\frac{1}{2}\ln\qty(1-\frac{2m}{r}) 
    \end{aligned}\label{single structure}        
\end{equation}
where $m$ is the mass enclosed by the radius $r$, $\Phi$ and $\Lambda$ are the metric functions, $p$ and $\varepsilon$ are the pressure and energy density respectively. 

In the two-fluid model the TOV equations become \cite{arpan_das}
\begin{equation}
    \begin{aligned}
        \frac{\dd m_1}{\dd r} &= 4\pi r^2 \varepsilon_1\\[5pt]
        \frac{\dd m_2}{\dd r} &= 4\pi r^2 \varepsilon_2\\[5pt]
        \frac{\dd \Phi}{\dd r} &=\frac{(m_1+m_2)+4 \pi r^3(p_1+p_2)}{r (r- 2(m_1+m_2))}\\[5pt]
        \frac{\dd p_1}{\dd r} &=-(p_1+\varepsilon_1)\frac{\dd \Phi}{\dd r}\\[5pt]
        \frac{\dd p_2}{\dd r} &=-(p_2+\varepsilon_2)\frac{\dd \Phi}{\dd r}\\[5pt]
        \Lambda (r)&= -\frac{1}{2}\ln\left\{1-\frac{2(m_1+m_2)}{r}\right\}
    \end{aligned} \label{eq: structure eqns twof}        
\end{equation}

These are now five coupled differential equations that need to be solved simultaneously, where the subscripts $1,2$ label the two fluids. At some point during the integration of these equations, the pressure of one of the fluids will vanish, marking the radius $R_i$ of fluid $i$. The integration is then switched to solve the single fluid-structure equations \eqref{single structure}. When the pressure of the remaining fluid vanishes, we get the final stellar structure. We get the total mass of the star as the sum of the masses of the two fluids. When the radius of the DM is less than that of the other fluid, the star has a DM core, while if the radius is greater, the star has a DM halo surrounding it. 

\subsection{Dimensionless tidal deformability}

A compact star in a binary experiences deformation due to the external quadrupolar field of the companion. Tidal deformability parameter is the measure of the response in terms of the deformability of the star to the external quadrupolar field and defined as the ratio of the induced mass quadrupole moment $Q$ to external perturbing tidal field $\cal{E}$ as \cite{Hinderer_tidal,kwing_leung}

\begin{equation}
    \lambda= -\frac{Q}{\cal{E}} = \frac 23 k_2R^5
\end{equation}
where
\begin{align}
    \nonumber k_2={}& \frac{8}{5}C^5(1-2C^2)\qty[2-y_R+2C(y_R-1)]\times\left\{2C(6-3y_R+3C(5y_R-8))\right.\\
    \nonumber &+4C^3\qty[13-11y_R+C(3y_R-2)+2C^2(1+yR)]\\
    &\left.+3(1-2C^2)\qty[2-y_R+2C(y_R-1)]\ln(1-2C)\right\}^{-1}.
\end{align}

$C=M/R$ is the compactness parameter with $M$ and $R$ being the mass and radius of the star respectively. Here $y$ is the solution to the differential equation
\begin{equation}
    y' = -\frac{1}{r}\qty(y^2+ ye^{2\Lambda}\qty(1+ 4\pi r^2(p- \varepsilon))+ Qr^2)    
\end{equation}

 where 

 \begin{equation}
     Q= 4\pi e^{2\Lambda}\qty(5\varepsilon+ 9p+ (p+\varepsilon)\dv{\varepsilon}{p})- \frac{6e^{2\Lambda}}{r^2}- 4\Phi'^2
 \end{equation}
 
 and $y_R$ is the value of $y$ at the surface of the star. From here, the dimensionless tidal deformability is defined as
\begin{equation}
    \Lambda = \frac{\lambda}{M^5} = \frac 23 \frac{k_2}{C^5}
\end{equation}
And the binary tidal deformability $\tilde{\Lambda}$ is
\begin{align}
\tilde{\Lambda} = \frac{16}{13} \frac{(12q + 1) \Lambda_1 + (12 + q) q^4 \Lambda_2}{(1 + q)^5},
\end{align}
here, $q$ is the mass ratio of the secondary component to the primary component. For two-fluid, DM admixed stars, the modifications in these equations are \cite{arpan_das}: 
\begin{align}
    \varepsilon\to \varepsilon_1+\varepsilon_2\qquad p\to p_1+p_2\qquad(p+\varepsilon)\dv{\varepsilon}{p}\to (p_1+\varepsilon_1)\dv{\varepsilon_1}{p_1}+ (p_2+\varepsilon_2)\dv{\varepsilon_2}{p_2}
\end{align}

\subsection{Dependence on DM parameters}\label{sec: DM param TOV}

With this two-fluid model, to get the mass-radius (M-R) relation of the star, first, we need to fix the DM fraction present in the star. We fix the DM fraction by mass as
\begin{equation}
\begin{aligned}
M_f = \frac{M(DM)}{M(DM)+M(HM)}
\end{aligned}
\end{equation}
Here M(DM) represents the total mass of the DM in the star, and M(HM) means the same for HM. We use DDME2 density-dependent parametrization for HM in this work \cite{2005PhRvC..71b4312L}. The mass fraction of DM in a particular star is determined completely by the central energy densities of the HM and DM. Since this fraction cannot be known until the structure equations are solved, we employ a root-finding algorithm to find the DM's central energy density, which would give the required mass fraction for a given central energy density of HM. As the two matter components do not interact with each other, two different spheres are formed. Depending on the central density and EOS of DM, the NS can have a DM core surrounded by HM, or have a DM halo surrounding an HM star. In the case of the M-R relations, we take the radius of the star to be the observable radius of the star (radius of HM) because the electromagnetic probes used to find the radius of the star cannot detect the DM halo \citep{rafiei2023tidal,2024PhRvD.109d3029S}. However, when calculating the tidal deformability, we take the star's radius to the maximum radius among the radius of HM and DM. This is because the DM halo would have a gravitational influence affecting the spacetime geometry \cite{rafiei2023tidal}.
\begin{figure}[h!]
    \centering
    \begin{subfigure}[b]{0.49\textwidth}
        \centering
        \includegraphics[width=\textwidth]{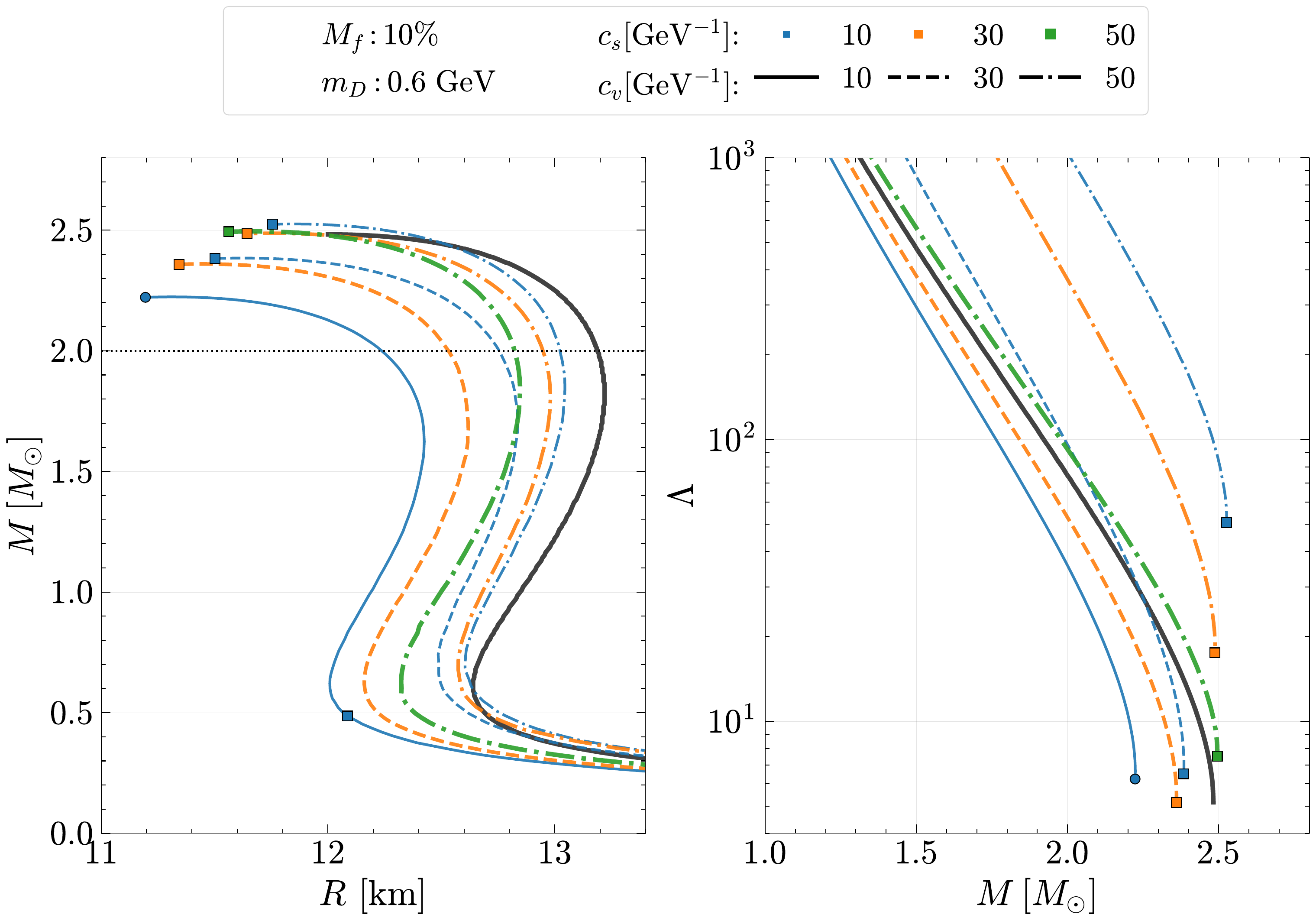}
        \begin{subfigure}[b]{\textwidth}
         \centering
         \includegraphics[width=1.025\textwidth]{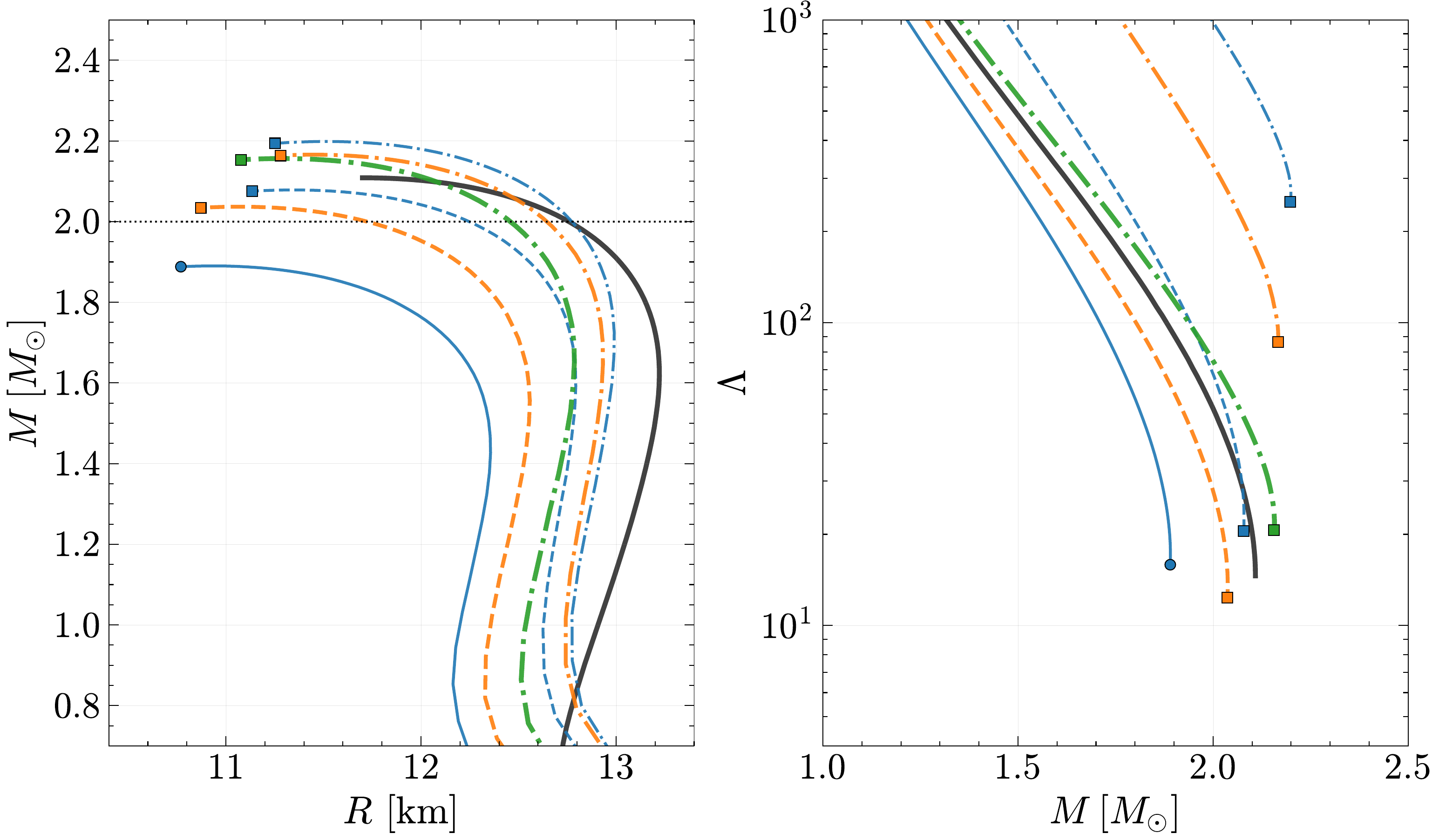}
         \end{subfigure}
    \end{subfigure}
    \begin{subfigure}[b]{0.49\textwidth}
        \centering
        \includegraphics[width=\textwidth]{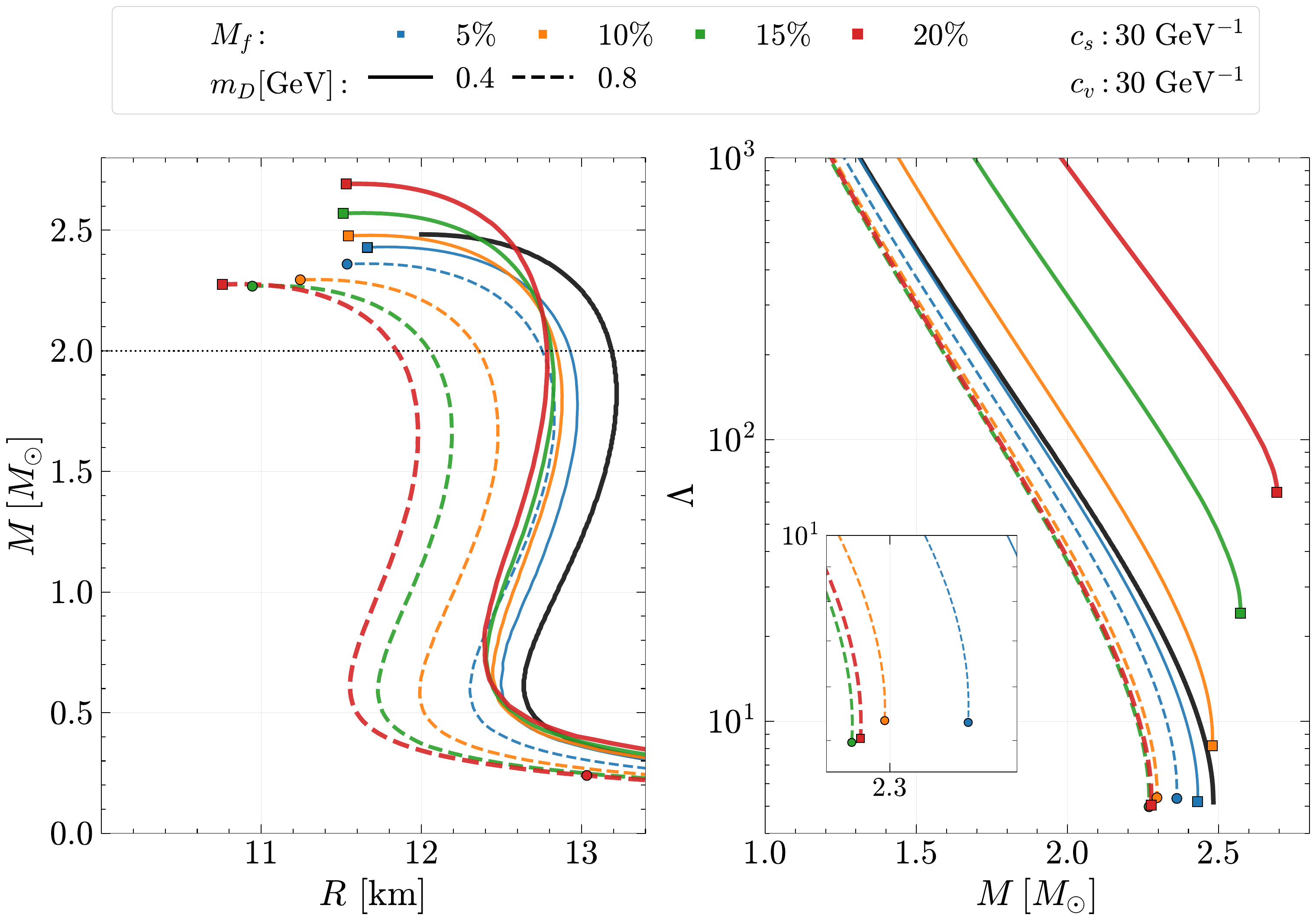}
        \begin{subfigure}[b]{\textwidth}
        \centering
        \includegraphics[width=1.025\textwidth]{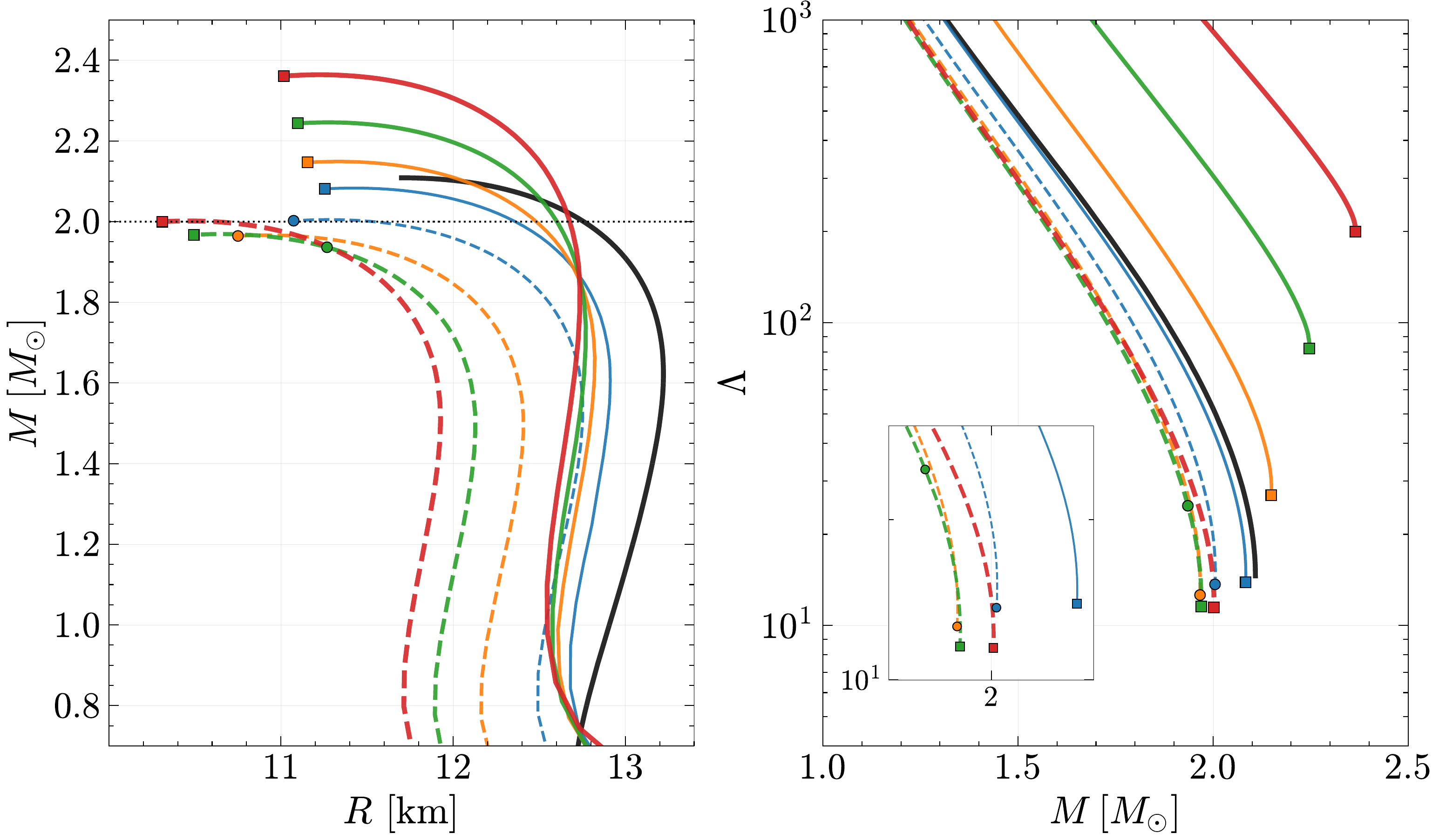}
        \end{subfigure} 
    \end{subfigure}
    \caption{(colors are available in the online version) Effects of varying the DM parameters on the M-R and tidal deformability with stellar mass relations of DM admixed stars. The upper panels are for pure nuclear matter, and the lower panels also include hyperons. The thick, solid black line is for pure BM. While decreasing mass, the square marker shows the conversion of DM core to halo and vice versa for the circle marker. The horizontal line in the M-R graph is for $2M_\odot$.}
    \label{fig:MR_tidal}
\end{figure}

\begin{figure}[h!]
    \centering
    \begin{subfigure}[b]{0.5\textwidth} 
    \centering
    \includegraphics[width=\textwidth]{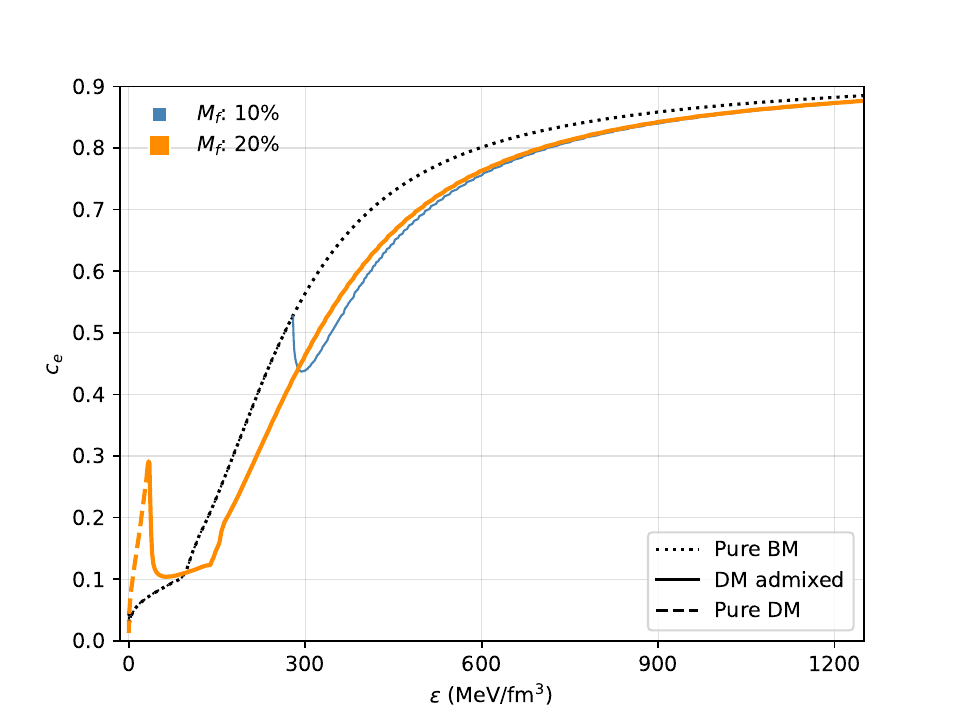}
    \end{subfigure} 
    \hspace{-0.75cm}
    \begin{subfigure}[b]{0.5\textwidth} 
    \centering
    \includegraphics[width=\textwidth]{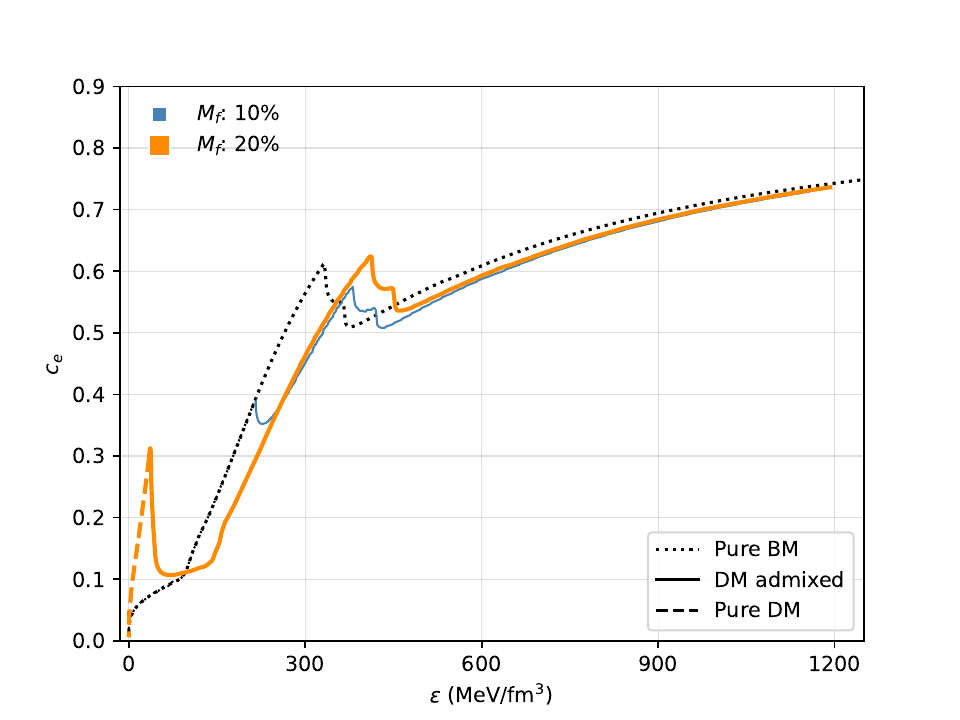}
    \end{subfigure}
    \caption{(colors are available in the online version) Equilibrium speed of sound plotted against average density and energy density in the star for maximum mass configuration. The left panel is for pure nuclear matter, and the right panel also includes hyperons. The value of both parameters $c_s$ and $c_v$ is 30 $\mathrm{GeV}^{-1}$ and dark matter particle mass is 0.8 $\mathrm{GeV}$.}
    \label{fig:speed_sound}
\end{figure}

For the EOS of the crust, we use the Baym-Pethick-Sutherland (BPS) model as given in reference \cite{1971ApJ...170..299B}. To observe the effects of different parameters on M-R relations and tidal deformability, we choose a few values from the set of parameters as mentioned in section \ref{sec: DM params}. For a fixed fraction $M_f =  10\%$ of DM with particle mass $m_D = 0.6$ $\mathrm{GeV}$, the effect of scalar and vector interactions in the dark sector is shown in the left panels of figure \ref{fig:MR_tidal}. In the upper panels, we show pure nucleonic matter, and in the lower panels, hyperons are also included. For obvious reasons, in the presence of hyperons, the EOS of matter becomes soft, reducing the maximum attainable mass and tidal deformability, as evident from a comparison of upper and lower panels. Here, the square marker shows the transition from DM core to DM halo as mass decreases and vice versa for the transition with the circle marker. For example, for $c_s=c_v=10\ \mathrm{GeV}^{-1}$, $m_D=0.6\ \mathrm{GeV}$ and $M_f=10\%$, every star with a mass above the square marker and below the circle marker has a DM core. Keeping the same vector interaction ($c_v$), with increasing scalar interaction ($c_s$) DM EOS becomes softer. Similarly, keeping $c_s$ the same,  DM EOS becomes stiffer with an increase of $c_v$. Stars having stiffer DM EOS would have DM distributed widely, making them less compact. That is why the DM halo is formed for these stars within the mass range as shown. This less dense DM causes a larger radius of the NS as compared to the NSs having DM cores due to dense DM; this is evident from the left panel of figure \ref{fig:MR_tidal}. In the sub-right panels, we show the variation of $\Lambda$ with mass for different values of $c_s$ and $c_v$. Again, for stiffer DM EOS, the tidal deformability is large as compactness is small. The horizontal dotted line represents the minimum $2~M_\odot$ maximum mass bound from the M-R relation.
\begin{figure}[h!]
    \centering
    \begin{subfigure}[b]{\textwidth}
    \centering
    \includegraphics[width=\textwidth]{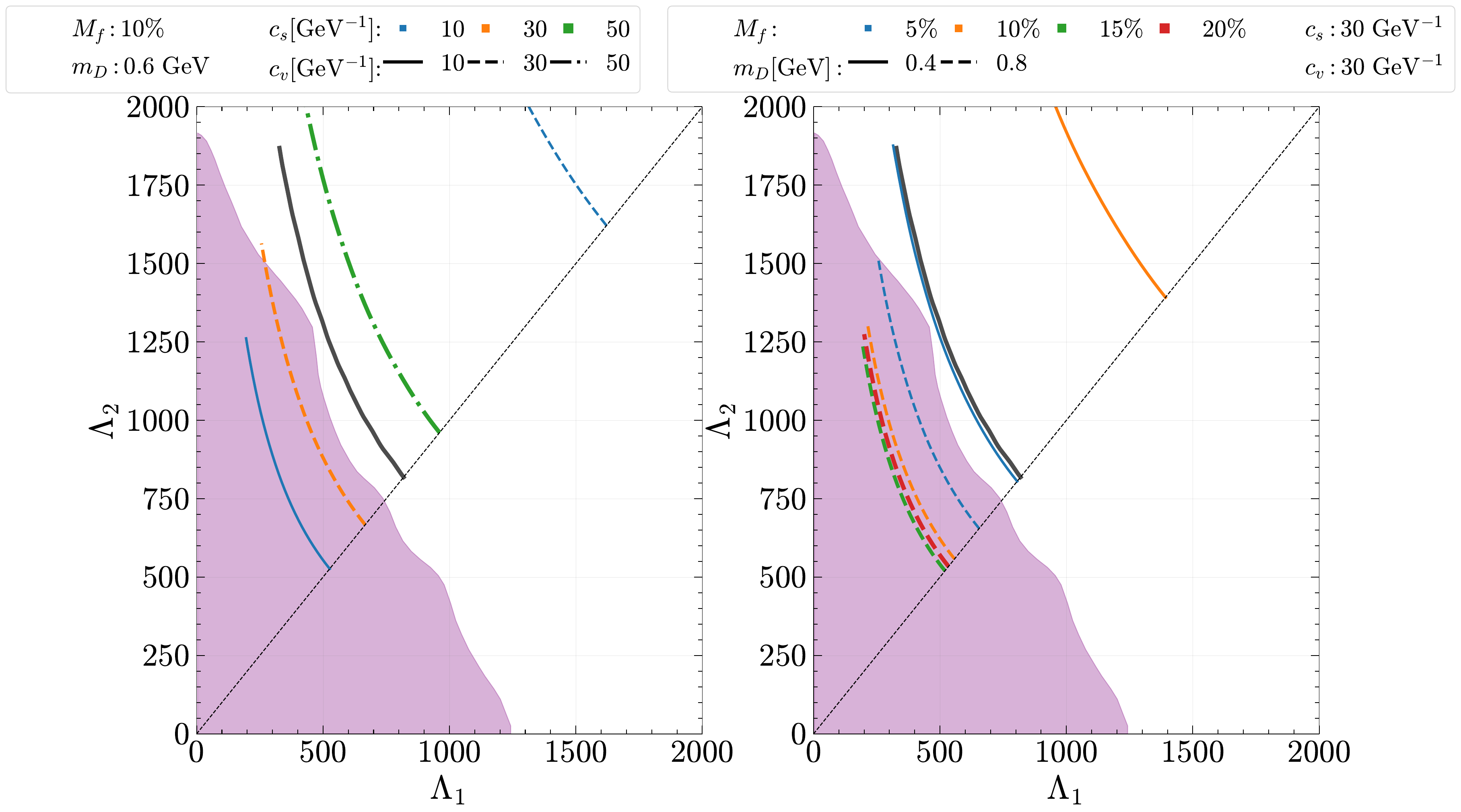}
    \end{subfigure}
    \begin{subfigure}[b]{\textwidth}
    \centering
    \includegraphics[width=0.875\textwidth]{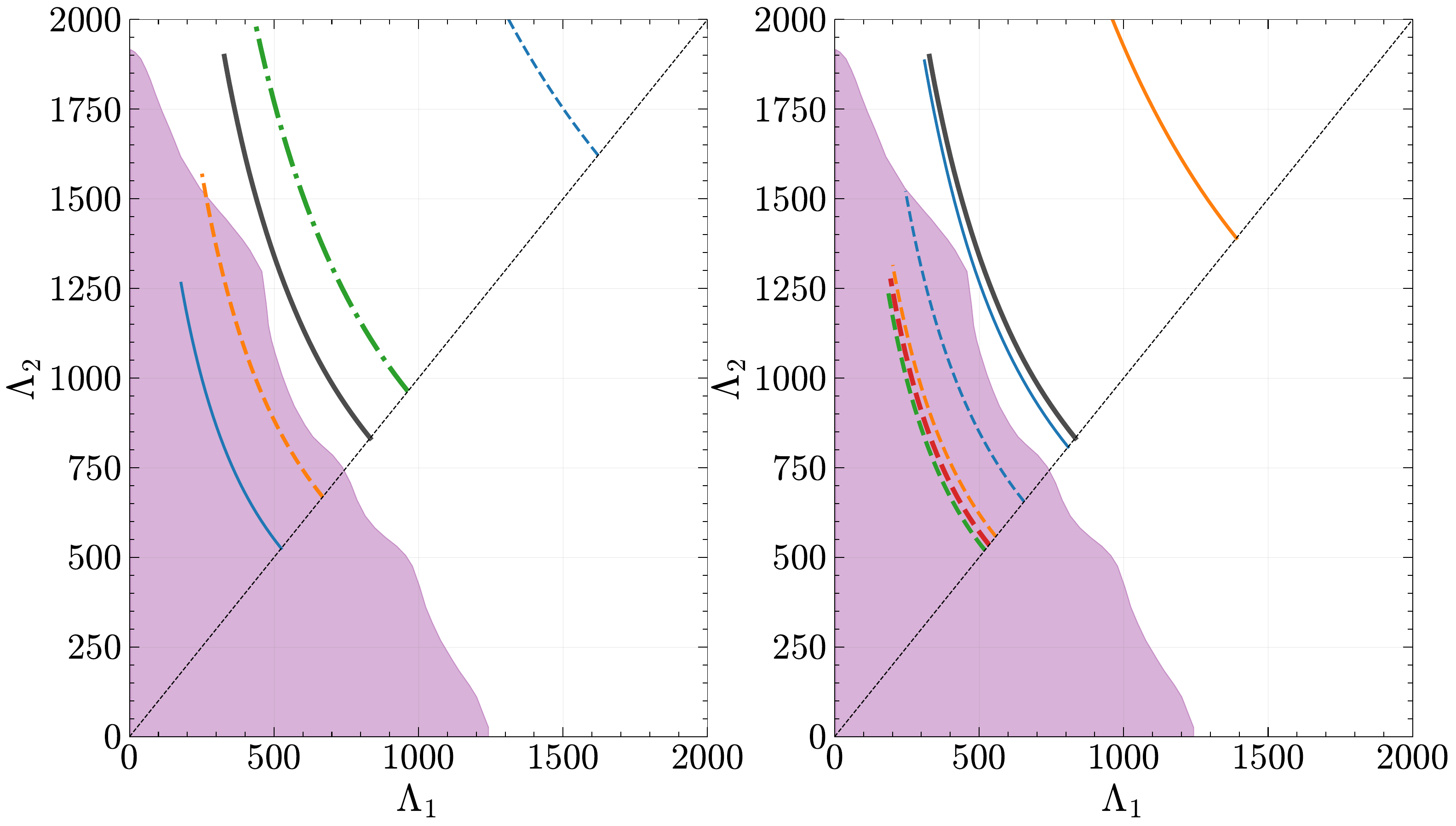}
    \end{subfigure}
    \caption{(colors are available in the online version) Effects of varying the DM parameters on the combinations of tidal deformability of primary and secondary mass in DM admixed stars. The upper panels are for pure nuclear matter, and the lower panels also include hyperons. The thick, solid black line is for pure BM.}
    \label{fig:binary_tidal}
\end{figure}
For the fixed scalar and vector interactions, with $c_s = c_v = 30$ $\mathrm{GeV}^{-1}$ the variation in M-R relation for different $M_f$ are shown in the right panel of figure \ref{fig:MR_tidal} by solid curves for $m_D = 0.4$ $\mathrm{GeV}$ and by dashed curves for $m_D = 0.8 \mathrm{GeV}$. Here, we observe that for larger $m_D$, the DM EOS becomes softer, mostly forming DM cores, and for smaller $m_D$ only DM halos are formed for the given combinations. For the softest DM EOS ($c_s=c_v=30\ \mathrm{GeV}^{-1}$, $m_D=0.6\ \mathrm{GeV}$ and $M_f=10\%$), we observe the transition from DM core to halo while stellar mass is increasing. In other cases with smaller $M_f$ a DM core is formed. In cases of DM halo, to get a larger $M_f$, the quantity of DM needs to increase. This increases the maximum mass of the star. In cases of DM core, lesser quantities of HM are needed for greater $M_f$. As a result, the maximum mass of the star decreases. This behavior is also evident from the variation of $\Lambda$, which is shown in the sub-right panel of figure \ref{fig:MR_tidal}. We observe that for smaller $m_D$ when DM halo is formed, increasing $M_f$ decreases the compactness, increasing $\Lambda$; while, for larger $m_D$ and cases of DM cores, increasing $M_f$ increases compactness, decreasing $\Lambda$. DM halos are more likely to be formed for lower $m_D$ and higher $M_f$. These conclusions are similar to those obtained by ref. \cite{liu_wei_DM_effects_2024}.
By fixing chirp mass and considering all the possible mass combinations such that the sum of both masses range is within the total mass range of GW170817 event, the tidal deformability of primary and secondary components is shown in figure \ref{fig:binary_tidal}. The shaded region represents the $\tilde{\Lambda}$ less than 720 as given by the reference \cite{PhysRevX.9.011001}. Curves with softer EOSs like lower $c_v$, higher $c_s$, and higher values of $M_f$ and $m_D$ lie within the limit. Results remain almost the same with the appearance of hyperons due to a very small fraction of hyperons at this energy density. The equilibrium speed of sound ($c_e = \sqrt{dp/d\varepsilon}$) is depicted in figure \ref{fig:speed_sound} with energy density for the maximum mass configuration corresponding to each set of parametrization. $c_e$ is within the range $0\leq{c_e}\leq{c}$ obtained from causality and thermodynamic conditions \cite{PhysRevLett.32.324}. 
As we noticed in previous results, with the same parameters, the maximum mass star with 10\% DM fraction has a DM core, and with 20\% DM fraction, a DM halo surrounds it. In the left panel, there are some sudden drops in the graph due to the appearance of another fluid; for example, the dashed curve for 20\% DM (orange) converts into a solid curve when HM gets mixed with DM, there is a drop due to softer EOS of HM with respect to DM EOS. In the right panel, an additional drop can be noticed at an energy density of approximately 400 MeV fm$^{-3}$ due to the appearance of hyperons with nucleons. The appearance of hyperons makes EOS soft, or it reduces the slope $dp/d\varepsilon$ suddenly \cite{2002PhRvL..89q1101S,2012NuPhA.881...62W}. 
The presence of DM significantly affects the speed of sound.

\section{Non Radial Modes}
\subsection{Equations governing non-radial modes}
Due to external or internal perturbations in the star, non-radial oscillations are generated. Here we concentrate on the quadrupolar ($l=2$) f- and p1-mode. These modes have pressure as the restoring force and radial node numbers 0 and 1, respectively. We calculate the mode frequencies with the relativistic Cowling approximation. Although the numerical values of frequencies in the Cowling approximation differ from those in the full GR calculations (by up to $30\%$ for $f\text{-}$, and \(\sim \)$15\%$ for $p_1\text{-}$ \cite{Kunjipurayil_f_p,Zhao_universal_f_g}), the qualitative results are not affected much.

We follow the formalism given in \cite{Sotani_Cowling,Thorne_Campolattaro}. Since we're working within the relativistic Cowling approximation, the metric perturbations are neglected. The fluid perturbations are governed by the fluid Lagrangian displacement vector, which is taken to be 
\begin{equation}
    \xi^i= \qty(\frac{e^{-\Lambda}}{r^2}W(t,r)\  Y_{lm}(\theta, \phi),\ \frac{-V(t,r)}{r^2} \ \partial_\theta Y_{lm}(\theta, \phi), \ \frac{-V(t,r)}{r^2\sin^2\theta}\  \partial_\phi Y_{lm}(\theta, \phi))\label{xi}
\end{equation}

Here $W$ and $V$ are variables of the fluid perturbation. Their relations are obtained by taking a variation of the energy-momentum conservation law $\delta(\nabla_\nu T^{\mu\nu})$ which reduces to $\nabla_\nu\delta T^{\mu\nu}$ in the Cowling approximation. Here $\delta T^{\mu\nu}$ is given by \cite{Thorne_Campolattaro}

\begin{equation}
    \delta T^{\mu\nu}= \begin{bmatrix}
        \ -\delta\varepsilon & \qty(p+\varepsilon)U^0 U^1 & \qty(p+\varepsilon)U^0 U^2 & \qty(p+\varepsilon)U^0 U^3 \ \\[10pt]
        \ \qty(p+\varepsilon)U^1 U^0 & \delta p & 0 & 0 \ \\[10pt]
        \ \qty(p+\varepsilon)U^2 U^0 & 0 & \delta p & 0 \ \\[10pt]
        \ \qty(p+\varepsilon)U^3 U^0 & 0 & 0 & \delta p\ 
    \end{bmatrix}\label{pert T}
\end{equation}

$U^\mu$ is obtained by simply taking a temporal derivative of equation \eqref{xi} and noting that the time-like component is the same as that of the unperturbed four velocity for a static perfect fluid in the metric given by equation \eqref{metric}: ${U}^0= e^{-\Phi}$

\begin{equation}
    U^\mu= \qty(e^{-\Phi},\ \frac{e^{-\Lambda}}{r^2}\partial_t W\ Y_{lm},\ \frac{-\partial_t V}{r^2} \ \partial_\theta Y_{lm}, \ \frac{-\partial_tV}{r^2\sin^2\theta}\  \partial_\phi Y_{lm})
\end{equation}

The Eulerian variations of energy density and pressure are given by \cite{Thorne_Campolattaro}
\begin{equation}
    \begin{aligned}
        \delta \varepsilon&= \qty(p+\varepsilon)\frac{\Delta n}{n} - \varepsilon'\xi^r\qquad
        \delta p&= \gamma p\frac{\Delta n}{n} - p'\xi^r
    \end{aligned}
\end{equation}

where
\begin{equation}
    \frac{\Delta n}{n}= -\qty(e^{-\Lambda}\frac{W'}{r^2}+\frac{l(l+1)}{r^2}V)Y_{lm}\label{dn_n}
\end{equation}

is the lagrangian variation in number density, and 
\begin{equation}
    \gamma= \pdv{\ln p}{\ln n}=\frac{p+\varepsilon}{p}\frac{\partial p}{\partial \varepsilon}\label{gamma}
\end{equation}
is the adiabatic compressibility index. The dash represents the derivative with respect to $r$. $W$ and $V$ are assumed to have a harmonic time dependence, i.e., $W(t,r)= W(r)e^{i\omega t}$ and $V(t,r)= V(r)e^{i\omega t}$. We now start framing the perturbation equations following the procedure given in \cite{Sotani_Cowling}. We substitute the perturbed variables and four-velocity in equation \eqref{pert T}, and take the covariant divergence of the perturbed energy momentum tensor for two free indices $\mu=r, \theta$. Throughout the derivation, we use $$p'= -\Phi'(p+\varepsilon)\qquad\text{and}\qquad\frac{p'}{p+\varepsilon}= -\Phi'$$ both of which are valid in the two-fluid case, with the singular terms replaced by the total ones, since from equation \eqref{eq: structure eqns twof}:

\begin{align}
    \nonumber p_1'+p_2'&= -\Phi'(p_1+\varepsilon_1+p_2+\varepsilon_2)= -\Phi'(p+\varepsilon)\ \ \text{and,}\\[1.5ex]
    \frac{p_1'+p_2'}{p_1+\varepsilon_1+p_2+\varepsilon_2}&=\frac{-(p_1+\varepsilon_1)\Phi'-(p_2+\varepsilon_2)\Phi'}{p_1+\varepsilon_1+p_2+\varepsilon_2}= -\Phi' \label{eq: pdash}
\end{align} 

To remove the dependence of $\gamma$, we use $$ \frac{p'}{\gamma p}= \frac{p'}{p+\varepsilon}\ \pdv{\varepsilon}{p}\ = -\Phi'\ \pdv{\varepsilon}{p}$$ which is easily obtained from equation \eqref{gamma}. Making all these substitutions, we finally get:

\begin{align}
    W'&= \pdv{\varepsilon}{p}\qty[\omega^2 r^2e^{\Lambda-2\Phi} V+\Phi'W]- e^\Lambda l(l+1)V\label{eq: dW}\\
    V'&= 2\Phi'V-\frac{e^{\Lambda}}{r^2}{W}+\qty(\dv{\varepsilon}{p}-\pdv{\varepsilon}{p})\qty[\Phi'\ V+\frac{e^{-\Lambda+2\Phi}}{\omega^2 r^2}{\Phi'^2}\ W]\label{eq: dV}
\end{align}

These are the oscillation mode equations. As a check of their correctness, note that if we take $\dv{\varepsilon}{p}= \pdv{\varepsilon}{p}$, these equations reduce to those of \cite{Sotani_Cowling}

\begin{equation}
    \begin{aligned}
        W'&= \dv{\varepsilon}{p}\qty[\omega^2 r^2e^{\Lambda-2\Phi} V+\Phi'W]- e^\Lambda l(l+1)V\\
        V'&= 2\Phi'V-\frac{e^{\Lambda}}{r^2}{W}\label{dW_dV}
    \end{aligned}
\end{equation}

Moreover, for stars with a temperature or composition gradient, these equations are the same as eqs. (25) and (26) of \cite{Zhao_Jaik_quasinormal_g} with the transformations $\nu\to2\Phi$, $\lambda\to 2\Lambda$ for the line element, and
\begin{equation*}
    \mathbb{W}\to -r^{-(l+1)}W\qquad \text{and}\qquad \mathbb{U}\to r^{-l} e^{-2\Phi} V
\end{equation*}
where $\mathbb{W}$ and $\mathbb{U}$ are the fluid perturbation variables in \cite{Zhao_Jaik_quasinormal_g} and $W$ and $V$ are the corresponding variables in this work.

The boundary conditions at the center of the star are $W(r)= Ar^{l+1}$ and $V(r)=-Ar^l/l$, where $A$ is an arbitrary constant. The surface boundary condition ($r=R$) is \cite{Sotani_Cowling} $\Delta p= \gamma p \frac{\Delta n}{n}= 0$ which from equation \eqref{dn_n} gives
\begin{equation}
    \omega^2 R^2e^{\Lambda(R)-2\Phi(R)} V(R)+\Phi'|_{r=R} W(R)= 0 \label{surf cond}
\end{equation}

In the two-fluid case, the only modification is in the $ \dv{\varepsilon}{p}$ term. Using equation \eqref{eq: pdash}, we get 
\begin{align}
    \dv{\varepsilon}{p}&= \dv{\qty(\varepsilon_1+\varepsilon_2)}{\qty(p_1+p_2)}\ =\  \dv{\varepsilon_1}{p_1}\frac{p_1'}{p_1'+p_2'}+ \dv{\varepsilon_2}{p_2}\frac{p_2'}{p_1'+p_2'}\\[1.5ex]
    \dv{\varepsilon}{p}&= \frac{p_1+\varepsilon_1}{p_1+\varepsilon_1+p_2+\varepsilon_2} \ \dv{\varepsilon_1}{p_1}+ \frac{p_2+\varepsilon_2}{p_1+\varepsilon_1+p_2+\varepsilon_2}\ \dv{\varepsilon_2}{p_2}
\end{align}

Since we don't consider temperature or composition gradients, the adiabatic sound speed equal to the equilibrium sound speed. Thus, equation \eqref{dW_dV}, along with the above boundary conditions, are the equations used in this work. Once the central energy densities for HM and DM for a particular mass fraction have been identified, we solve the oscillation mode equations to find the frequency $\omega$ subject to boundary conditions. We start with an initial guess for $\omega$ and integrate equation \eqref{dW_dV}. We then check if that value of $\omega$ satisfies equation \eqref{surf cond}. If not, the guess of $\omega$ is improved via the Newton-Raphson algorithm. The lowest root of equation \eqref{dW_dV} which satisfies the boundary conditions, $\omega$, is the frequency corresponding to the $f\text{-}$mode. The next highest root is the first ($p_1\text{-}$mode) in a family of pressure modes with ever-increasing frequency and radial node numbers. We check the radial node number to classify the modes by counting the number of times $W$ and $V$ are trivial inside the star.

\subsection{Dependence on DM parameters and the hyperons presence}

\begin{figure}[h!]
    \centering
    \begin{subfigure}[b]{0.49\textwidth}
        \centering
        \includegraphics[width=\textwidth]{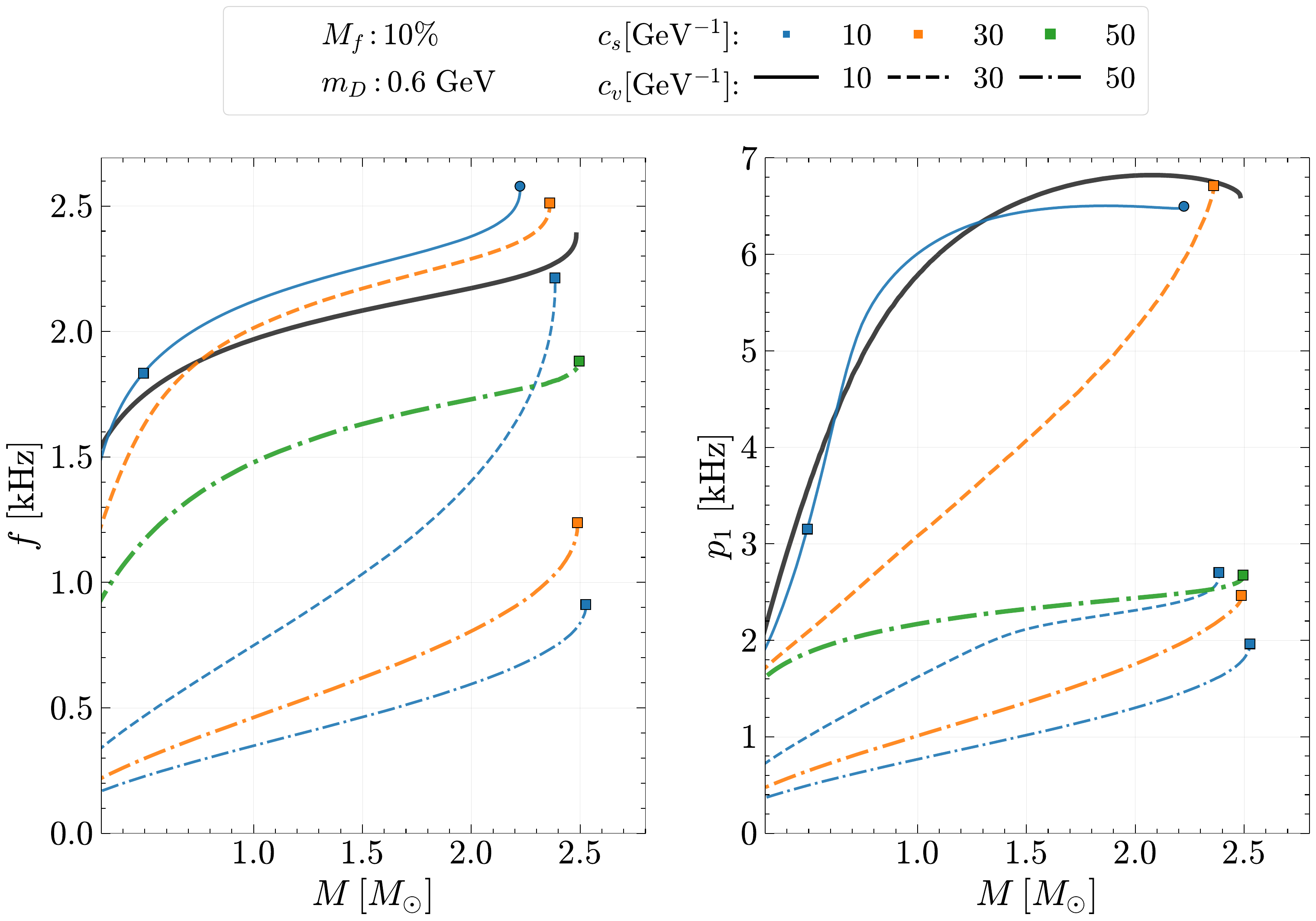}
        \begin{subfigure}[b]{\textwidth}
         \centering
         \includegraphics[width=1.025\textwidth]{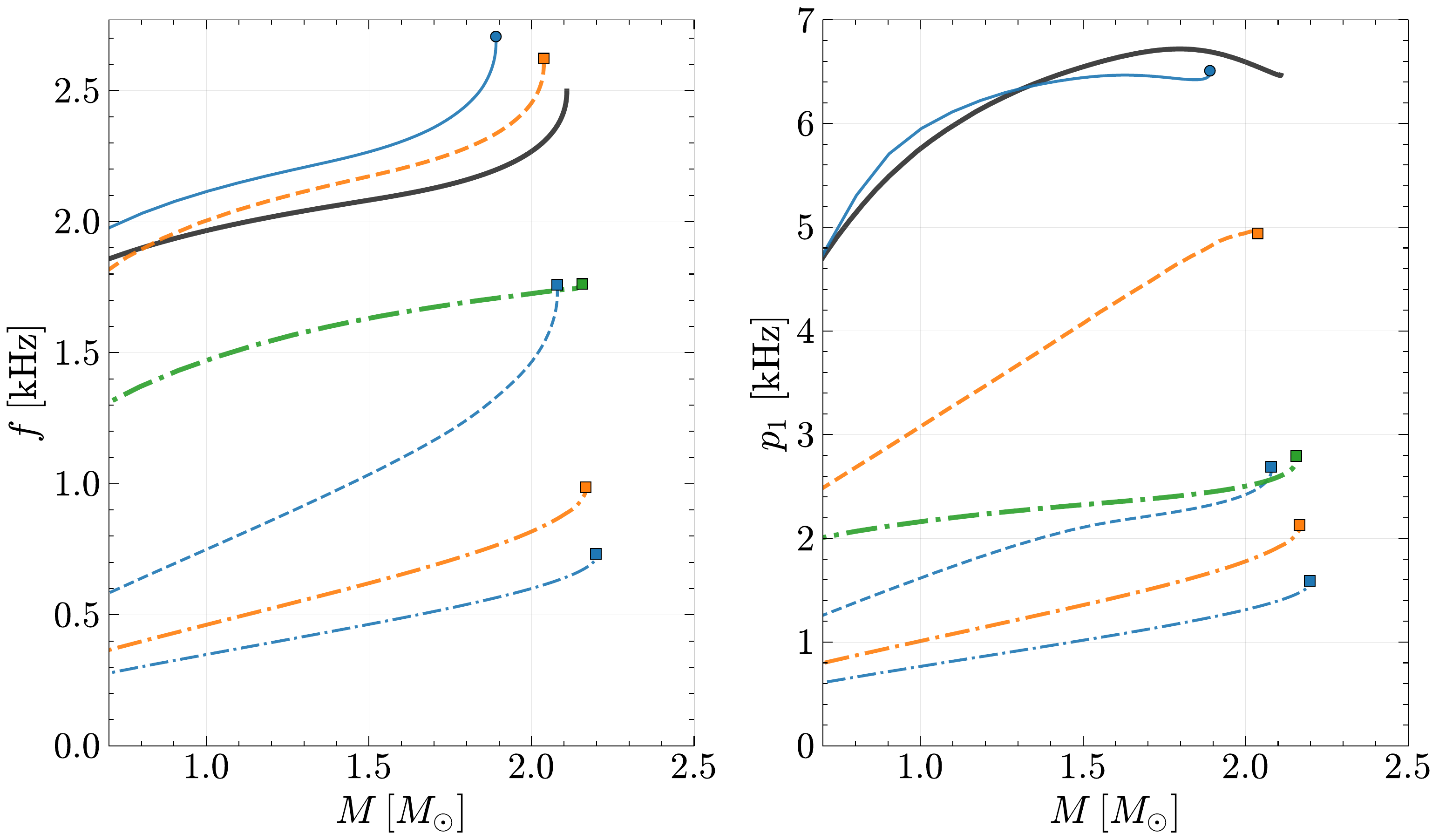}
         \end{subfigure}
    \end{subfigure}
    \begin{subfigure}[b]{0.49\textwidth}
        \centering
        \includegraphics[width=\textwidth]{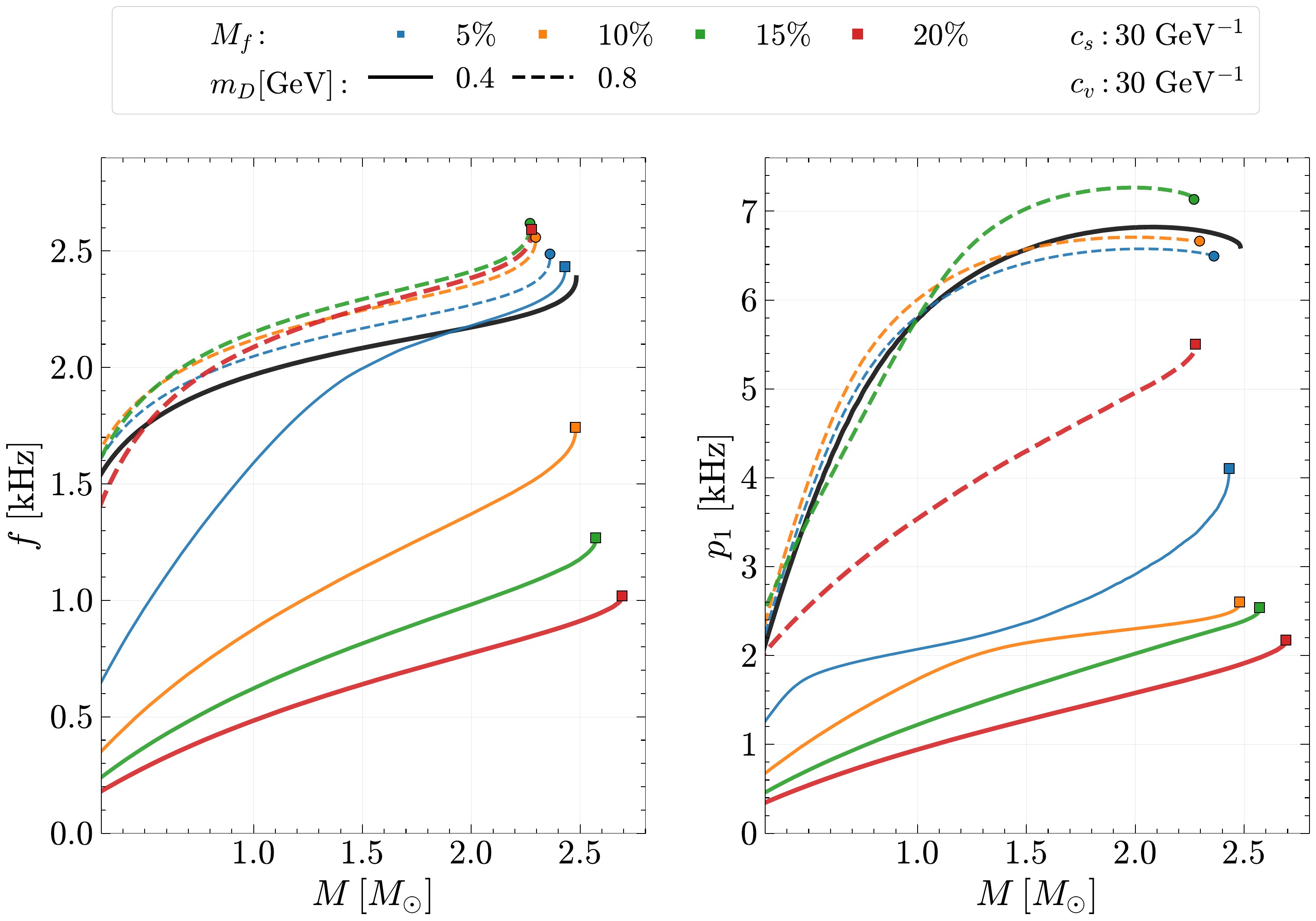}
        \begin{subfigure}[b]{\textwidth}
        \centering
        \includegraphics[width=1.01\textwidth]{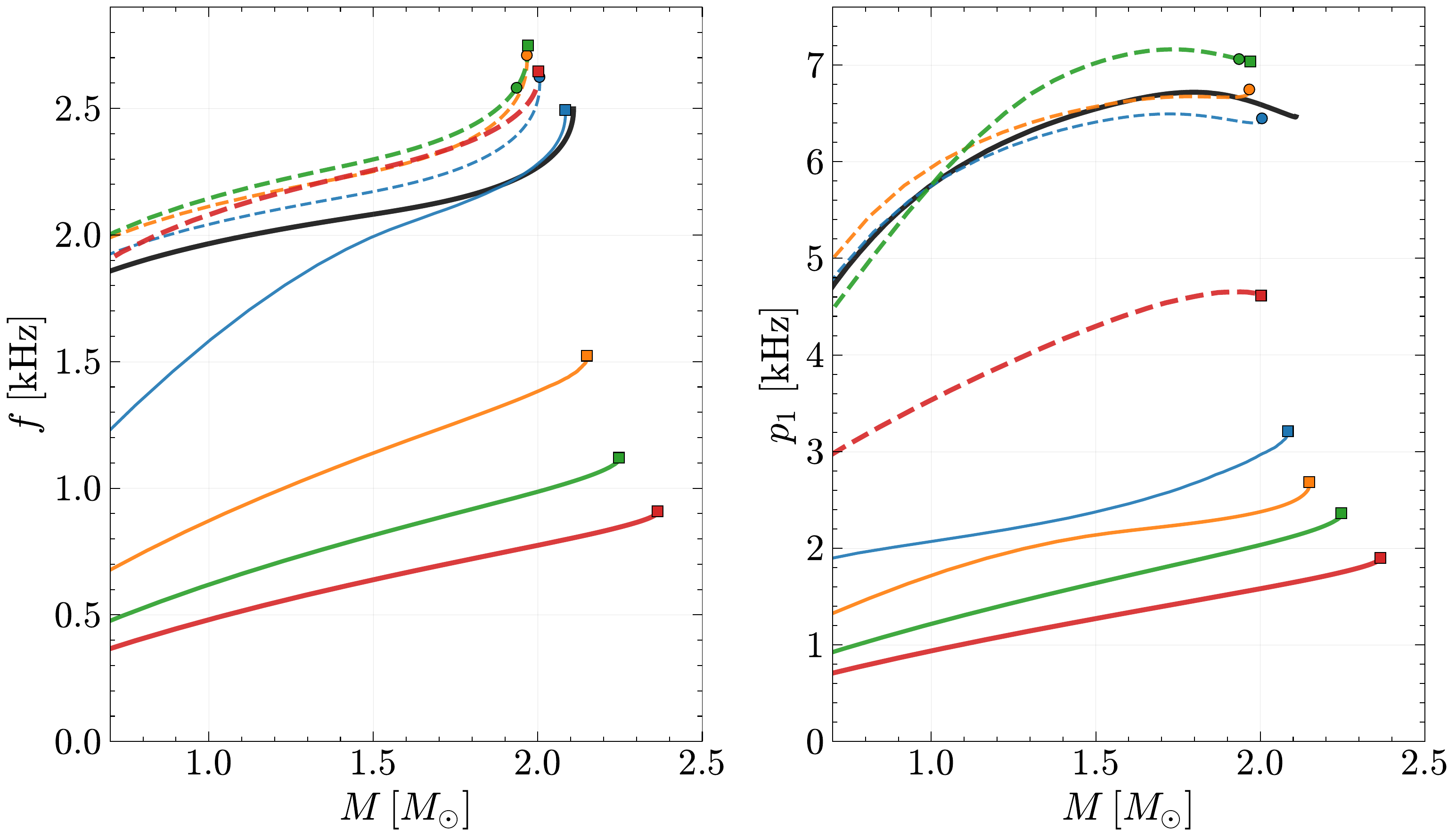}
        \end{subfigure} 
    \end{subfigure}
    \caption{(colors are available in the online version) Effects of varying the DM parameters on the $f$ and $p_1$ mode frequency with stellar mass in DM admixed stars. The upper panels are for pure nuclear matter, and the lower panels for matter with hyperons. The thick, solid black line is for pure BM. While decreasing mass, the square marker shows the conversion of DM core to halo and vice versa for the circle marker.}
    \label{fig:fnp_mode}
\end{figure}

 For the same DM parameters as in section \ref{sec: DM param TOV}, we see the effects on the non-radial mode frequencies. Figure \ref{fig:fnp_mode} shows the variation of $f\text{-}$ and $p_1\text{-}$modes frequencies with stellar mass for different parametrizations and $M_f$ of DM. As we have seen larger values of $c_s$ for a fixed $c_v$ make the matter in the star softer, the $f\text{-}$ and $p_1\text{-}$mode frequency increases while larger values of $c_v$ for a fixed $c_s$ have the opposite effect. This is clear from the left panels of the figure \ref{fig:fnp_mode}. The effect on $p_1\text{-}$mode frequencies is most notable since the frequencies can decrease a lot, even becoming comparable to the typical ranges of the $f\text{-}$mode frequency (1-3 kHz) by suitably varying the DM self-interaction strength.
\begin{figure}[t!]
    \centering
    \includegraphics[width=\linewidth]{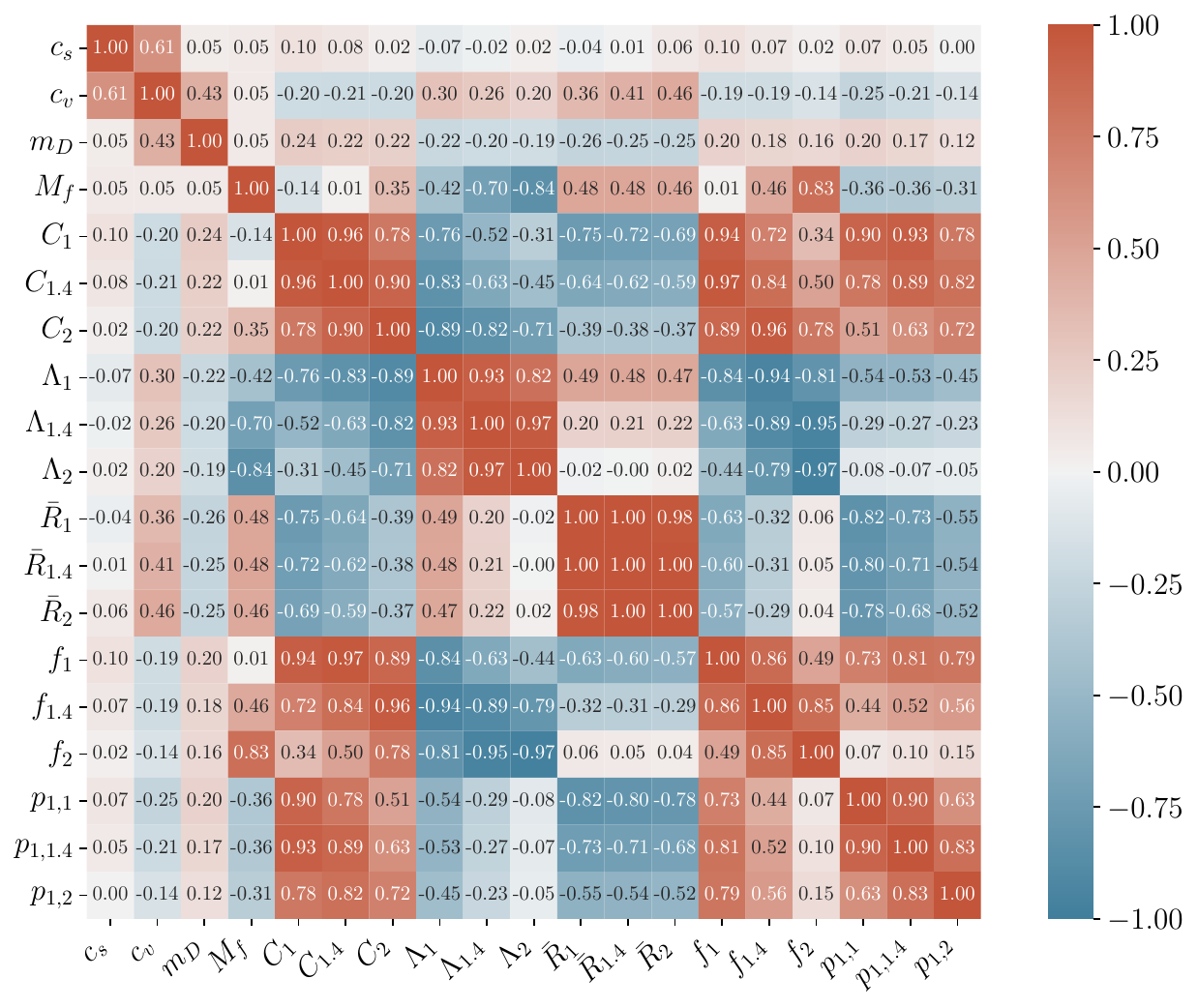}
    \caption{(colors are available in the online version) Correlation matrix showing the correlations amongst the DM EOS parameters, stellar structure parameters, and the non-radial modes. Correlations are obtained after applying the minimum 2$M_\odot$ maximum mass and $\tilde{\Lambda}\leq720$ constraints. Hyperons are not included in HM EOS.}
    \label{fig:correlations}
\end{figure}
In the right panels of figure \ref{fig:fnp_mode}, we show the variation of non-radial mode frequencies with stellar mass for different $m_D$ and $M_f$. For both the $f\text{-}$ and $p_1\text{-}$mode cases the variation of frequencies with DM mass fraction is sensitive to whether the star has a DM halo or core. The mode frequencies are less for the DM halo and more for the DM core. When the stars have a DM halo, increasing $M_f$ for a fixed $m_d$ decreases the mode frequencies as it increases the stiffness, while for stars with DM core, the opposite effect is observed. Again, the effect on the $p_1\text{-}$mode is much stronger than the effects on the $f\text{-}$mode. This is because $p_1\text{-}$modes, due to the presence of a radial node, are much more sensitive to the distribution of matter in the star \cite{anderssonGravitationalWaveAsteroseismology1998}. Presence of hyperons shows effects on non-radial oscillation mode's frequency near the maximum mass. In this region, if the DM core exists inside the star, then the frequency of both modes increases more rapidly than the DM halo.
\begin{figure}[h!]
    \centering
    \begin{subfigure}[b]{0.65\textwidth}
       \includegraphics[width=\linewidth]{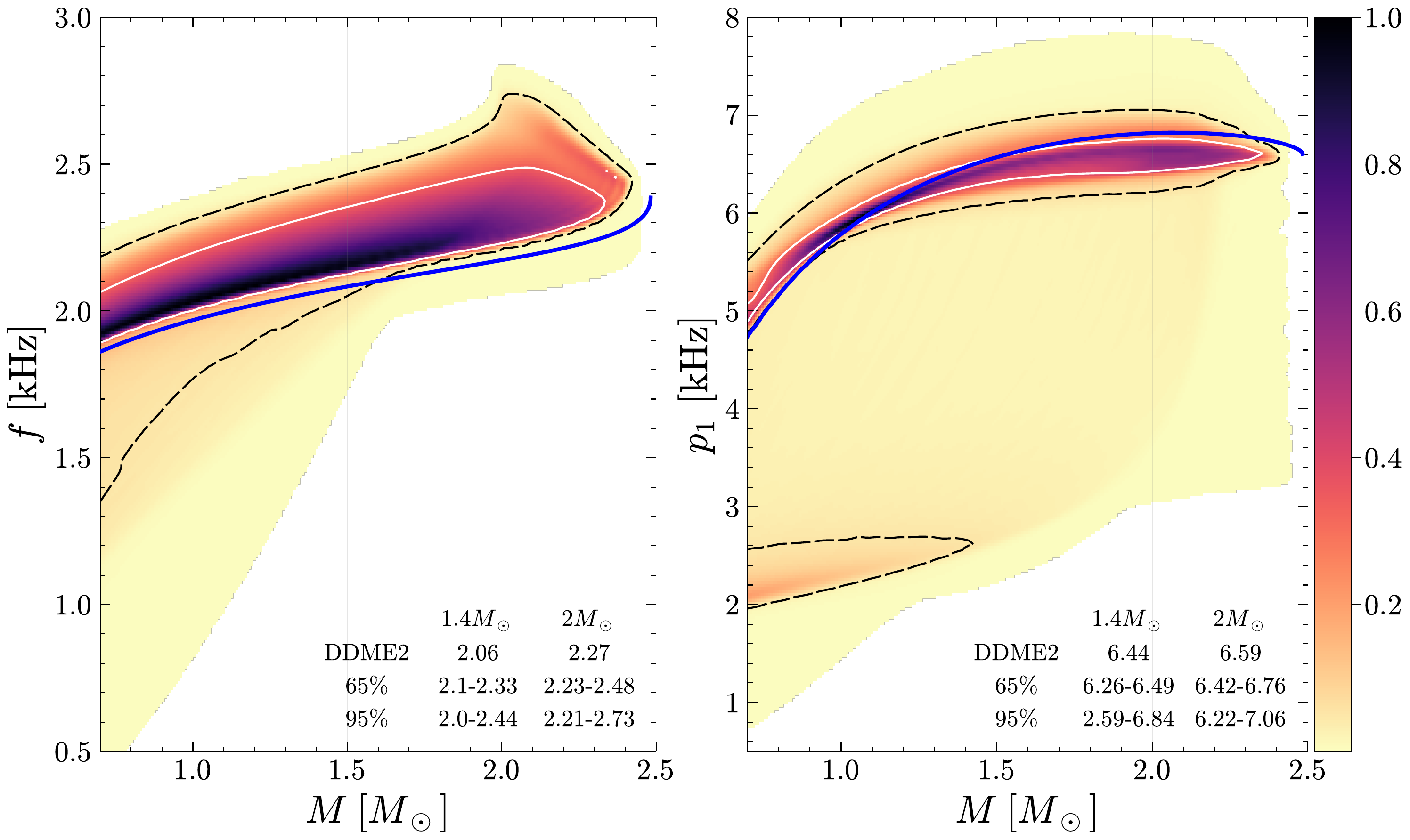} 
    \end{subfigure}
    \vspace{1cm}
    \begin{subfigure}[b]{0.65\textwidth}
       \includegraphics[width=\linewidth]{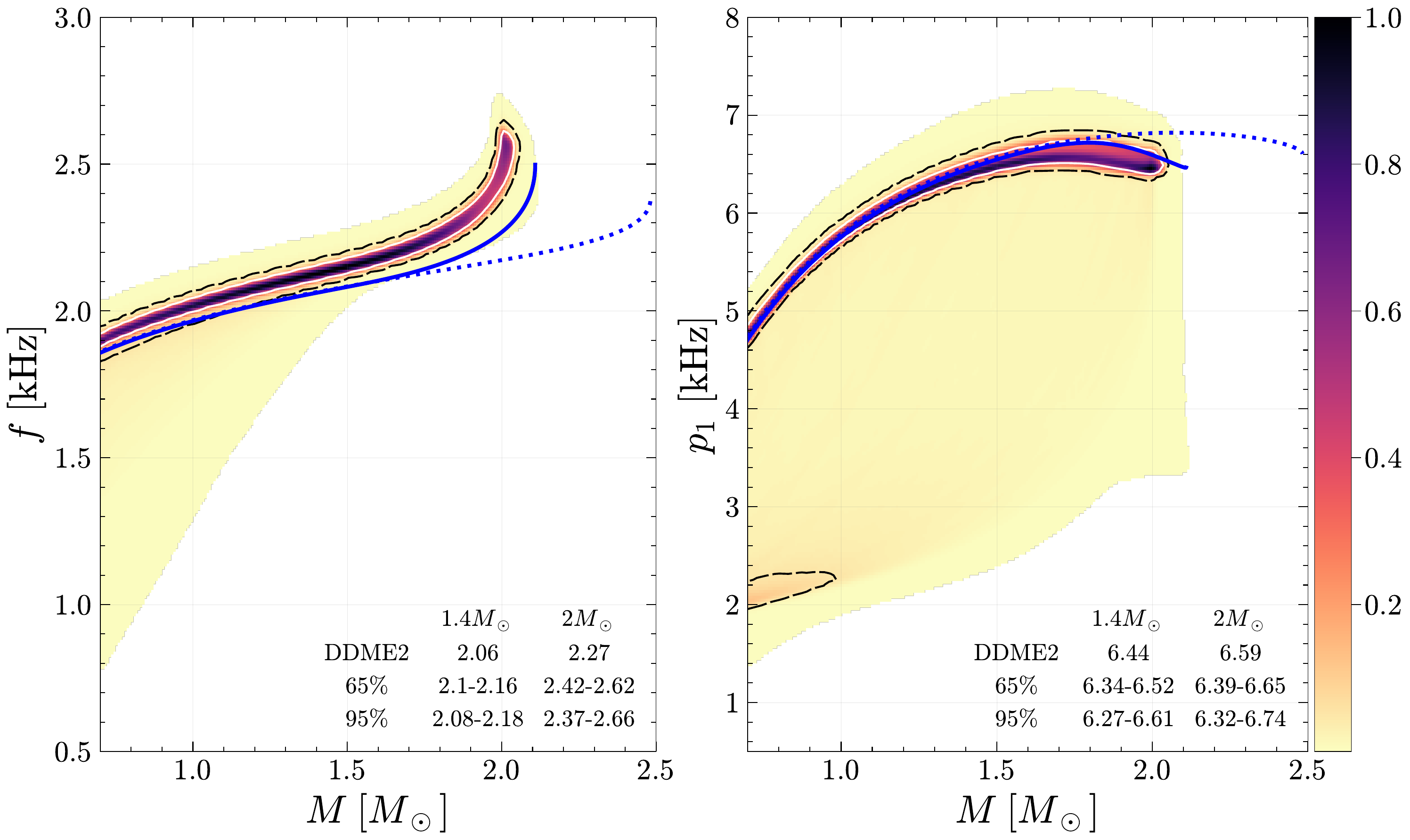}  
    \end{subfigure}
    \caption{(colors are available in online version) PDFs for $f\text{-}M$ and $p_1\text{-}M$ relations for valid stars. The upper panels are for pure nuclear matter, and the lower panels also include hyperons. The thick, solid blue line is for pure BM. In lower panels, the blue dotted line represents pure nuclear matter for comparison. The white solid and the black dashed lines represent the $65\%$ and $95\%$ regions, respectively. The $f\text{-}M$ and $p_1\text{-}M$ relations for pure HM (DDME2) are shown with a solid blue line. Frequency values given by the confidence intervals at 1.4$M_\odot$ and 2$M_\odot$ are mentioned along with the frequency values for DDME2 at the same masses.}
    \label{fig:prob_fnp}
\end{figure}
\subsection{Correlation studies}
We show the mutual dependence of these parameters by the correlations between them in figure \ref{fig:correlations}. Since the effects of DM are heavily dependent on the combinations of the parameters $c_s$, $c_v$, $m_D$, and $M_f$, correlations with any one of these parameters is expected to be weak. However as expected, the mode frequencies are strongly correlated with the corresponding compactness. We define compactness as the total mass of the star ($M_{HM}+M_{DM}$) divided by the outermost radius of the star (radius of DM, $R_{DM}$ in case of DM halo, and radius of HM, $R_{HM}$ in case of DM core). The $p_1\text{-}$modes have a stronger negative correlation with another parameter $\bar{R}=R_{DM}/R_{HM}$ as compared to $f\text{-}$modes. This parameter signifies the distribution of DM with respect to HM. This means that the $p_1\text{-}$mode frequency is more sensitive to the presence of either a DM halo or a DM core than the $f\text{-}$mode. 
The behavior of the mode frequencies is very similar to that of the tidal deformability as shown in figure \ref{fig:MR_tidal}. When a particular variation of parameters increases the tidal deformability, a decrease in frequency is observed for the same variation. This suggests a negative correlation of the modes with tidal deformability and is indeed what we see in figure \ref{fig:correlations}. Tidal deformability also shows a strong negative correlation with compactness. 
\newpage
\begin{figure}[h!]
    \centering
    \begin{subfigure}[b]{0.55\textwidth}
        \includegraphics[width=\linewidth]{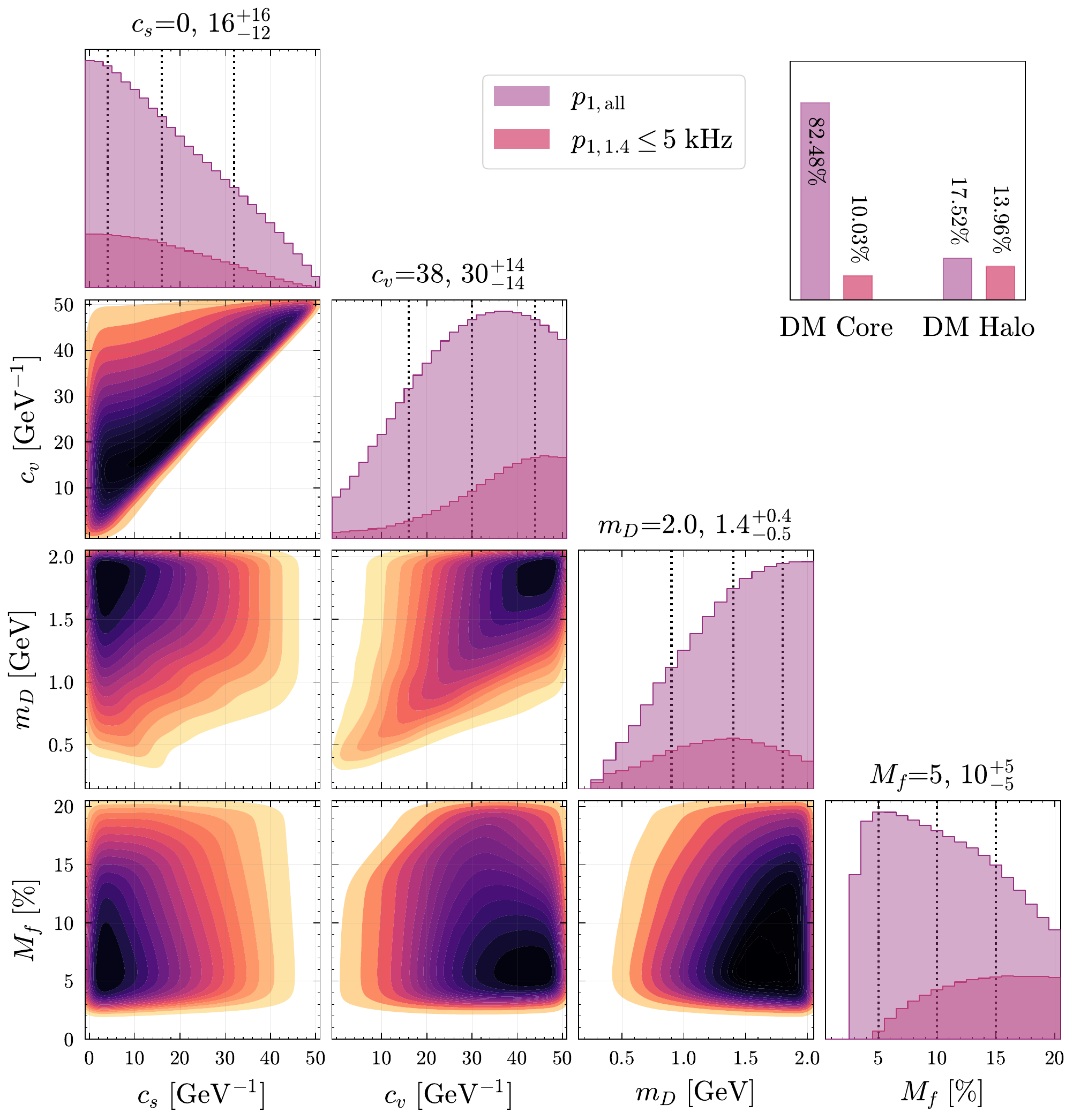}
    \end{subfigure}
    \begin{subfigure}[b]{0.55\textwidth}
        \includegraphics[width=\linewidth]{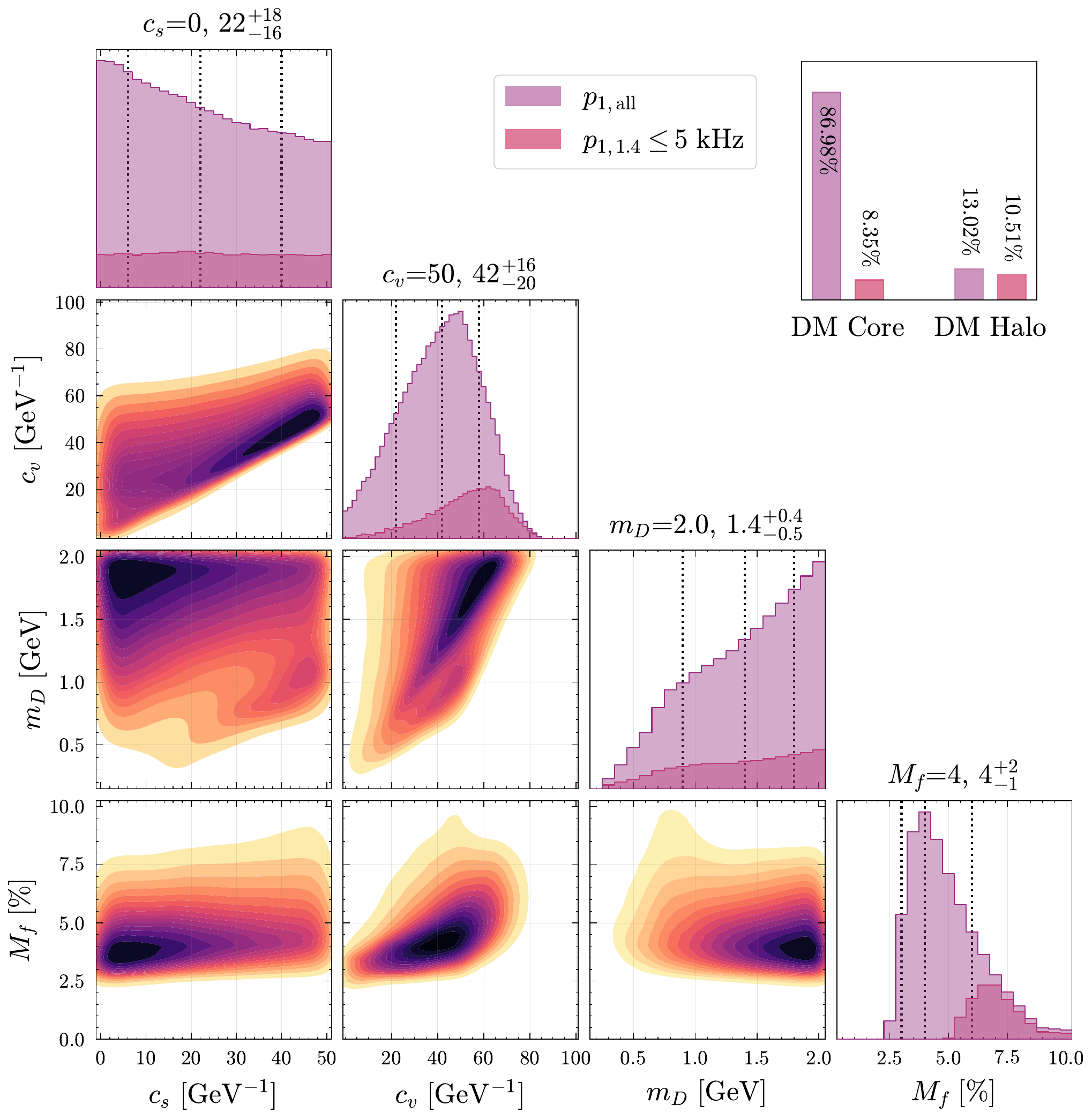}
    \end{subfigure}
    \caption{(colors are available in the online version) Corner plot with KDEs for the free parameters used in this work. The upper panel is for pure nuclear matter, and the lower panel is for nuclear matter with hyperons. The most probable values of each parameter, along with the median value with $1\sigma$ deviation, have been mentioned above the diagonal plots. The bar plot shows the percentage of stars with a DM core or halo. A darker color is used to mark the subset of low-frequency $p_1\text{-}$modes}
    \label{fig:prob}
\end{figure}
\newpage
\section{DM parameters from astrophysical observations}
Next, we constrain the DM parameters for the DM admixed NS from the astrophysical observations of compact objects. We generate M-R relations and tidal deformability with all possible combinations of DM EOS parameters ($c_s,~c_v,~m_D$) and $M_f$, as mentioned in section \ref{sec: DM params} and then filter our EOSs by making them satisfy the mass constraint of at least 2$M_\odot$ maximum mass and the tidal deformability constraint of $\tilde{\Lambda}\leq720$. We call DM-admixed stars following these constraints `valid'. Once we have all combinations of parameters that follow these observational constraints, we show a KDE corner plot in figure \ref{fig:prob}. From the $2,56,880$ total combinations, only $59,030$ combinations give valid DM EOSs with $M-R$ and $\Lambda-M$ curves which follow the observational constraints we impose. We plot probability density plots for the $f\text{-}$ and $p_1\text{-}$modes in figure \ref{fig:prob_fnp}. For this, we first make a $250\times250$ grid and count the number of lines passing through each grid cell. Upon normalizing, we then have a density map of the $f\text{-}M$ and $p_1\text{-}M$ plane. From this, the $65\%$ (white solid line) and $95\%$ (black dashed line) confidence intervals can be found. We have also shown the $f\text{-}$ and $p_1\text{-}$ modes for DDME2 with no DM as a blue solid line. We find that the $f\text{-}$mode frequency for most DM admixed stars takes on a higher value than pure HM. For low mass stars, the $f\text{-}$mode frequency decreases a lot, with a minimum of around $1.5\ \mathrm{kHz}$ at $1.4\ M_\odot$. However, the change in the $p_1\text{-}$mode frequency is much more significant. Although the $65\%$ confidence interval encloses the $p_1\text{-}$mode for pure HM, there is a distinct low-frequency region for the $95\%$ confidence interval. This frequency, between $1-3\ kHz$ is the typical frequency range of $f$-modes. The lowest $p_1\text{-}$mode frequency at $1.4M_\odot$ is around $2.1\ \mathrm{kHz}$. Further, due to the sensitivity of $p_1\text{-}$modes to the DM distribution, the spread in their frequencies is much larger than that of the $f\text{-}$modes. In lower panels, the inclusion of heavier baryons reduces the maximum mass of neutron stars, and the presence of DM further exacerbates this reduction. We know that softer EOSs result in higher oscillation frequencies. However, we exclude the very soft EOSs that would result in a maximum mass lower than 2 $M_\odot$, as this restriction influences the maximum frequency value. Consequently, we obtain very narrow 65\% and 95\% confidence intervals for the frequencies of both f- and $p_1$-modes. For a massive star with a mass of 2 $M_\odot$, the 65\% credible interval for the f-mode frequency ranges from 2.42 to 2.62 kHz, which is narrower than the range obtained with the pure nuclear matter EOS. The 95\% credible region for a 1.4 $M_\odot$ star is more confined, with the f-mode frequency ranging from 2.08 to 2.18 kHz. Additionally, the range of the $p_1$-mode frequency with a 95\% confidence interval narrows significantly to 6.27-6.61 kHz, compared to the previous range of 2.62-6.87 kHz for a 1.4 $M_\odot$ star. This indicates that even for low-mass stars, the $p_1$-mode frequency is high.

We find that the most probable ($1\sigma$) values for or DM parameters are $c_s=0\ (4-32)\ \mathrm{GeV}^{-1}$, $c_v=38\ (16-44)\ \mathrm{GeV}^{-1}$, $m_D=2.0\ (0.9-1.8)\ \mathrm{GeV}$ and $M_f=5\ (5-15)\%$. Due to presence of hyperons the parameters becomes $c_s=0\ (6-40)\ \mathrm{GeV}^{-1}$, $c_v=50\ (22-58)\ \mathrm{GeV}^{-1}$, $m_D=2.0\ (0.9-1.8)\ \mathrm{GeV}$ and $M_f=4\ (3-6)\%$. With hyperons, EOS becomes soft. Thus, higher values of $c_v$ are more favorable for making EOS stiffer. So, values of $c_v$ are considered maximum up to 100 $\mathrm{GeV}$. Since increasing $m_d$ already reduces the maximum attainable mass and softens the EOS, large $c_s$ values are less probable. However, large $c_v$ values are more probable since they stiffen the EOS, increasing the maximum mass. Here, it should be noted that the most probable values of the DM parameters depend on the HM EOS chosen. Softer HM EOSs will allow for stiffer DM EOSs and larger $c_s$ values, which follow the constraints we imposed. So the range of DM parameters given here is only relevant for DDME2 EOS. Further, since increasing $m_D$ softens the DM-admixed star while increasing $c_v$ stiffens it, we can get valid DM-admixed stars for very large values of $m_D$ and $c_v$. As an example, a combination of $c_s=0\ \mathrm{GeV}^{-1}$, $c_v=50\ \mathrm{GeV}^{-1}$, $m_D=50\ \mathrm{GeV}$ and $M_f=7\%$ gives a valid star. The maximum accretion of DM predicted in neutron stars is typically less than this mass fraction, as further accumulation could lead to gravitational instability and collapse into a black hole \cite{2020Univ....6..222D,2010PhRvD..81l3521D}. However, self-interacting DM with strong vector interactions considered in this study exhibits stiff EOS and is comparable to nuclear matter. This stiffness can counteract the gravitational collapse, thereby preventing the formation of a black hole. Consequently, NS may sustain significantly higher DM fractions, potentially allowing for the presence of a surrounding DM halo. In this work, we have restricted ourselves to small $m_D$ values.

The darker regions in the histogram correspond to the low frequency $p_1\text{-}$modes. We find that most of our DM-admixed stars have DM cores. However, for stars with DM halo, most of them also have low-frequency $p_1\text{-}$modes. This means that combinations that are more likely to form a DM halo, i.e. higher $c_v$, lower $m_d$, and large $M_f$ are more likely to give low-frequency $p_1\text{-}$modes.

\section{Summary}
We have studied the stellar structure and properties of non-radial oscillation frequencies of DM admixed NSs. We have considered the density-dependent RMF model for the normal HM of the NS and self-interacting but non-annihilating DM which interacts with HM only gravitationally. The DM admixed NS can naturally have a large range of masses and tidal deformabilities since the DM EOS is not constrained. On the other hand, the observed radius of compact objects is related to the extent of HM in the DM admixed NSs. Hence observed M-R relations of the compact objects can be used to constrain the theoretical models of DM admixed NS. Due to their extreme densities, NSs are likely to be excellent captors of particle DM \cite{bramanteDarkMatterCompact2023}. However, if we assume that the astrophysical compact objects contain some fraction of DM, we can estimate some DM properties from astrophysical observations. For example, in our present study, from the observed lower bound of maximum attainable mass of the compact objects and the upper bound of tidal deformability from the binary star merger observations, we predict the maximum probable value of the DM self-interaction parameters and the DM particle mass within a specific model of NS. We also estimate the DM fraction within a DM admixed NS. We find that $5\%$ is the most repeating value of the DM fraction parameter. The higher chemical potential of nucleons allows the presence of heavy baryons. For the comparison, we include hyperons and find the most probable value of DM fraction as 4\%. Such a high dark matter fraction exceeds the maximum limit predicted by Bondi accretion models. However, it may be achievable for Fermion-Proca stars \cite{2024Parti...7...52J,2023arXiv230812174J}, which we will try to explore in our future studies. 

The existence of DM admixed NS compels us to study the NS properties which are also important for future GW observations. In light of this, we study the non-radial oscillation frequencies of DM admixed NSs. M-R and tidal deformability constraints help us to filter relevant DM parameters, and we predict the non-radial oscillation frequencies of DM admixed NS with these filtered parameters. We find that the $f\text{-}$ and $p_1\text{-}$mode frequencies are strongly correlated with compactness and distribution of DM. The $f\text{-}$mode frequency increases due to the presence of DM as compared to pure HM stars. But a different trend is observed with the $p_1\text{-}$mode frequency (typically 4-7 kHz), as the range is more widespread, with frequencies reducing up to the typical range of $f\text{-}$mode frequencies (1-3 kHz). After including hyperons, the 95\% confidence interval range of $p_1\text{-}$ mode shrinks noticeably, and it becomes to 6-7 kHz for a 1.4 solar mass star. A majority of these low frequency $p_1\text{-}$modes arise in stars with a DM halo and are more probable in low mass stars. Further, the $p_1\text{-}$modes for low-mass stars are more sensitive to DM distribution (whether there is a DM core or DM halo) as compared to $f\text{-}$modes which is evident from our corner plots and correlation studies. An important conclusion of our paper is thus that the effect of DM capture by compact stars can be seen much more clearly from non-radial modes, especially the $p_1\text{-}$modes.

\section*{Data Availability}
The data used in the manuscript can be obtained at reasonable request from the corresponding author.

\section*{Acknowledgments}
The authors acknowledge the financial support from the Science and Engineering Research Board (SERB), Department of Science and Technology, Government of India through Project No. CRG/2022/000069. The authors thank Ritam Mallick for some valuable suggestions and Kamal Krishna Nath for his input. The authors also want to thank the anonymous referee for enhancing the manuscript's quality through constructive comments.

\bibliography{references}
\bibliographystyle{unsrt}
\end{document}